\documentstyle[emulateapj]{article}

\begin{document}

\oddsidemargin  -0.5pc
\evensidemargin -0.5pc

\slugcomment{Submitted to ApJ October 14, 1999; Accepted February 02, 2000}

%---------------start of defs.-----------------------------------------
%defs.tex
\def\abs#1{\left| #1 \right|}
\def\EE#1{\times 10^{#1}}
\def\gcm{\rm ~g~cm^{-3}}
\def\cm3{\rm ~cm^{-3}}
\def\kms{\rm ~km~s^{-1}}
\def\cms{\rm ~cm~s^{-1}}
\def\isotope#1#2{\hbox{${}^{#1}\rm#2$}}
\def\wl{~\lambda}
\def\wll{~\lambda\lambda}
\def\Ha{{\rm H}\alpha}
\def\Hb{{\rm H}\beta}
\def\Lya{{\rm Ly}\alpha}
\def\Vej{V_{\rm ej}}
\def\Msun{~M_\odot}
\def\no{\hang\noindent}
\def\dots{$\ldots$}
\def\etal{{\it et al.}}
\def\ie{{\it i.~e.\ }}
\def\Pdot{\dot P}

%---------------end of defs.-----------------------------------------
 
\title{OBSERVATIONS OF THE CRAB NEBULA AND ITS PULSAR IN THE FAR-ULTRAVIOLET 
AND IN THE OPTICAL
\altaffilmark{1,2}
}

\author{Jesper Sollerman\altaffilmark{3,4,5}, 
Peter Lundqvist\altaffilmark{3},
Don Lindler\altaffilmark{6},
Roger A. Chevalier\altaffilmark{7},
Claes Fransson\altaffilmark{3}
Theodore R. Gull\altaffilmark{6},
Chun S.J. Pun\altaffilmark{6},
George Sonneborn\altaffilmark{6}
}

\altaffiltext{1}{
Based on observations with the NASA/ESA {\it Hubble Space Telescope},
obtained at the Space Telescope Science Institute, which is operated by the
Association of Universities for Research in Astronomy, Inc. under NASA
contract No. NAS5-26555.}
\altaffiltext{2}{Based on observations obtained at the {\it Nordic Optical 
Telescope} on La Palma, using the Andalucia Focal Reducer and Spectrograph.}

\altaffiltext{3}{Stockholm Observatory, SE-133 36 Saltsj\"obaden, Sweden.} 
\altaffiltext{4}{European Southern Observatory, 
Karl-Schwarzschild-Strasse 2, D-857 48 Garching bei M\"unchen, Germany.}
\altaffiltext{5}{Send offprint requests to Jesper Sollerman; 
E-mail: jesper@astro.su.se}
\altaffiltext{6}{Goddard Space Flight Center, Code 681, Greenbelt, MD 20771} 
\altaffiltext{7}{Department of Astronomy, University of Virginia, P.O. 
Box 3818, Charlottesville, VA 22903}

\begin{abstract}
We present far-UV observations of the Crab nebula and its 
pulsar made with the Space Telescope Imaging Spectrograph onboard 
the {\it Hubble Space Telescope}. Broad, blueshifted absorption arising in the
nebula is seen in C~IV~$\lambda$1550, reaching a blueward velocity of 
 $\sim 2500 \kms$. This can be interpreted as evidence for a fast outer shell 
surrounding the Crab nebula, and we adopt a spherically symmetric model to
constrain the properties of such a shell. From the line profile we find that
the density appears to decrease outward in the shell. A likely lower limit 
to the shell mass is $\sim 0.3\Msun$ with an accompanying kinetic energy
of $\sim 1.5\EE{49}$~ergs. 
%This occurs for the density structure $\rho(R) \propto R^{-3}$. 
A fast massive shell with $10^{51}$~ergs 
cannot be excluded, but is less likely if the density profile is 
much steeper than $\rho(R) \propto R^{-4}$ and the maximum velocity 
is $\lesssim 6000 \kms$. The observations cover the 
region $1140-1720$ \AA, which is further into the ultraviolet 
than has previously been obtained for the pulsar. With the 
time-tag mode of the spectrograph we obtain the pulse profile in this 
spectral regime. The profile is similar to that previously obtained by
us in the near-UV, although the primary peak is marginally narrower.
Together with the near-UV data, and new optical data
from the {\it Nordic Optical Telescope}, our spectrum of the Crab pulsar 
covers the entire region from $1140 - 9250$~\AA. 
Dereddening the spectrum with a standard extinction curve we achieve 
a flat spectrum for the reddening parameters $E(B-V)=0.52$, $R=3.1$. This 
dereddened spectrum of the Crab pulsar can be fitted by a power law with 
spectral index $\alpha_{\nu} = 0.11\pm0.04$. The main uncertainty in 
determining the spectral index is the amount and characteristics of the 
interstellar reddening, and we have investigated the dependence of 
$\alpha_{\nu}$ on $E(B-V)$ and $R$.
%In the optical we do not see the absorption 
%feature reported by Nasuti et al. 
In the extended emission covered 
by our $25 \arcsec \times 0\farcs5$ slit in the far-UV, we detect 
C~IV~$\lambda$1550 and He~II~$\lambda$1640 emission lines from the Crab nebula.
Several interstellar absorption lines are detected along the line of sight 
to the pulsar. The Ly$\alpha$ absorption indicates a column density of
$(3.0\pm0.5)\EE{21}$ cm$^{-2}$ of neutral hydrogen, which agrees well
with our estimate of
$E(B-V)$=0.52 mag. Other lines show no evidence of severe depletion 
of metals in atomic gas. 
\end{abstract}


\section{Introduction}
%--------------------%
The Crab nebula and its pulsar (PSR 0531+21) are among the most studied
objects in the sky. The discovery of the Crab pulsar 
as a fast rotating radio pulsar
(\cite{SR68};~\cite{C69}) paved the way for the interpretation of pulsars as 
neutron stars (\cite{G68}). 
Also, the position of the Crab pulsar in the center of the Crab
nebula, 
which is the remnant of supernova 1054, clearly supports the supernova -
neutron star connection. Soon after the radio detection the pulsar was
also shown to emit optical pulsations (Cocke, Disney, \& Taylor
1969). This established
that the pulsating star was the well known south preceding star in
the center of the nebula, which early optical spectroscopy showed to
emit a featureless continuum (\cite{M42}). To date, more than 1000 
%press release http://www.jb.man.ac.uk/news/pr9803.html nov 98 was#1000
radio pulsars are known, but only the following few 
have optical counterparts known 
to pulsate also in visible light: 
the Crab pulsar (\cite{Co69}), the LMC pulsar 0540-69 (\cite{MP85}),
the Vela pulsar (\cite{Wa77}), PSR 0656+14 (\cite{Sh97}) and 
(possibly) the Geminga pulsar (\cite{Sh98}). In the near-UV (NUV),  
pulsations have only been established for the Crab pulsar 
(\cite{P93};~\cite{G98}, henceforth G98).
Due to the faintness of these objects in the optical and in the ultraviolet,
the spectroscopic information is very limited. PSR 0540-69 was observed with
the Faint Object Spectrograph (FOS) onboard {\it Hubble Space Telescope (HST)}
in the $2500-5000$~\AA\ range (\cite{H97}) and showed a rather steep 
power law spectrum. These observations were, however, contaminated by nebular 
emission. The Geminga pulsar was observed with the {\it Keck} telescope 
(Martin, Halpern, \& Schiminovich 1998), but
the spectrum has very 
low signal-to-noise because this pulsar is exceedingly faint in the optical.
The only pulsar for which good signal-to-noise spectroscopy in the 
optical and ultraviolet
can be obtained is the Crab pulsar. Surprisingly enough,
very little has been done in this respect
since the optical observations of
Oke (1969).
In particular, until
the study by G98, no UV spectroscopy of the Crab pulsar had been published
since the first attempts by the {\it International Ultraviolet Explorer
(IUE)} (\cite{Ben80}). The {\it IUE} data cover only the NUV region 
($2000-3150$~\AA) 
and have poor signal-to-noise. The {\it HST}/STIS (Space Telescope Imaging 
Spectrograph) data from G98, and the new data presented here, clearly 
supersede these early attempts.

The Crab pulsar has been extensively studied over a very broad wavelength 
range, from the radio up to $\gamma$-rays (e.g., \cite{LGS98}). 
The high energy emission, from infrared (IR) to $\gamma$-rays, is believed 
to be the result of the same emission mechanism (e.g., \cite{LGS98}).
It is therefore of interest to fill in the gaps in the observed spectrum of 
the pulsar in this range. Although the pulsar is relatively
bright in the optical, UV observations are difficult due to the large 
extinction toward the Crab, $E(B-V)\sim0.5$ mag 
(e.g., \cite{DF85}, and references
therein). Here, we present UV observations of the Crab pulsar further into the 
far-UV (FUV) ($1140-1720$ \AA) than has previously been obtained. 
These are presented together with our previous NUV-data ($1600-3200$ \AA)(G98)
and new optical data from the {\it Nordic Optical Telescope (NOT)}.
Due to the large extinction correction, 
great care must be taken to draw conclusions 
about the intrinsic spectrum, and thus the emission mechanism of 
the pulsar. However, this procedure might also give a hint on the absorption
properties of the dust in the direction toward the pulsar.

In addition to
the pulsar emission, we detect emission lines from the Crab nebula 
itself in the FUV. In particular, the strength of the C~IV~$\lambda$1550 
emission can be of interest for abundance determinations.
Even more interesting is the broad C~IV~$\lambda$1550 
absorption line from the nebula detected against the pulsar continuum.
This line provides information on the nature of the SN 1054 event.

Although the Crab nebula has 
been studied extensively, the nature of the progenitor remains unknown. 
According to models based on the existence of the central neutron
star, as well 
as on nucleosynthesis arguments, the zero-age main
sequence (ZAMS) mass of the progenitor was probably 
in the range $8-13 \Msun$ (Nomoto 1985). The amount of material observed 
in the nebula ($4.6\pm1.8 \Msun$) seems too low to account for this 
(Fesen, Shull, \& Hurford
1997). Furthermore, the velocities of the filaments ($\sim 1400 \kms$,
Davidson \& Fesen 1985) give an uncomfortably low kinetic energy 
($\lesssim 1\EE{50}$ ergs) 
compared to other supernova remnants, i.e., at least an order of
magnitude less 
than the canonical energy of supernovae, $10^{51}$ ergs. 

The ``missing mass'' could either be in a slow progenitor wind, or in a fast,
hitherto undetected, shell  ejected at the explosion (Chevalier 1977). 
If the latter is true, as is
hinted by the observations of the outer [O~III] skin of the nebula 
(Hester et al. 1996; Sankrit \& Hester 1997),
this shell might account for the missing mass and kinetic energy of
the nebula.
The question remains, however, why such a 
shell has escaped detection despite many efforts to observe it 
(see, e.g., \cite{Fe97}). One possibility is that the low density of
the surrounding gas is not high enough to give rise to detectable 
circumstellar emission when interacting with the ejecta; 
neither X-ray nor radio searches have 
indicated any evidence of circumstellar interaction 
between fast ejecta and ambient gas
(Mauche \& Gorenstein 1989; Frail et al. 1995).
If a fast shell is absent, the birth of the Crab was definitely a low energy
event. This would call for a revision of our understanding of 
supernova explosions, especially since SN 1054 apparently was not unusually 
dim according to historical records (Chevalier 1977; Wheeler 1978). 

It is thus of great interest to further investigate whether there is a stellar 
wind or supernova ejecta outside the observed nebula, and what velocity this 
gas may have.  Lundqvist, Fransson, \& Chevalier 
(1986, henceforth LFC86) proposed to search for a 
fast shell by looking in the UV toward the Crab pulsar. Their time
dependent photoionization calculations showed that 
C~IV~$\lambda$1550 could show up in blueshifted absorption if 
the ionization history of the shell was as predicted in some models of 
Reynolds \& Chevalier (1984). Here, we present the detection of this
broad absorption line, and discuss its implications for the fast shell
around the Crab nebula.

First we discuss the observations and reductions (\S 2). We then 
(\S 3) discuss the pulse profile, the amount of reddening toward
the pulsar and the intrinsic pulsar spectrum in the optical/UV. In \S 3 we
also discuss the lines originating from the interstellar gas toward
the pulsar and from the Crab nebula itself. 
Some of these observations constrain the properties of a possible outer shell.
In \S 4 we summarize our conclusions.

%--------------------------------%
\section{Observations, Reductions and Results}
\subsection{{\it HST} Far-UV Observations}

The Crab pulsar was observed on January 22, 1999 using {\it HST}/STIS 
(\cite{K98}) with the FUV Multi Anode Micro-channel Array (MAMA) detector. 
The low resolution grating G140L was used,
which covers the wavelength interval $1140 - 1720$ \AA. These observations
were made in the time-tag mode and used a slit of $52\arcsec \times 0\farcs5$. 
The spectral resolution is 0.58 \AA\ pixel$^{-1}$, and the plate scale
is $0\farcs0244$ pixel$^{-1}$. This means that only 25\arcsec\ of 
the long-slit is actually projected onto the detector.
In total, six orbits of observations, including target acquisition, were used. 
These were divided into two visits. A log of
the observations is shown in Table 1. The total on-target exposure
time amounted to 14,040 seconds.
The orientation of the slit is shown in Figure 1.

\subsubsection{Time-resolved emission}

The time-tag mode on the STIS allows us to resolve the emission from
the Crab pulsar both in wavelength and time. The time resolution
obtained in this mode is 125 $\mu$s.
As these observations were the first to utilize the time-tag
capabilities of {\it HST}/STIS in the FUV for a known periodic variable,
special software, developed at Goddard Space Flight Center (GSFC) was used 
to obtain the pulse profile for the  pulsar.
The analysis followed the procedures outlined in G98. 
For each of the six datasets, a time averaged image was produced to trace 
the position of the pulsar spectrum. A 13 pixel wide window was 
used to extract events in 
the pulsar spectrum as well as in the background emission at
both sides of the pulsar emission. 
The arrival time of each 125 $\mu$s sample was converted to a solar system
barycenter arrival time. 
%The difference in the {\it HST} and solar system barycenter arrival time 
%is the dot product of the combined {\it HST} position w.r.t. the solar
%system barycenter and the unit vector in the direction of the crab pulsar
%divided by the velocity of light. 
The position of the {\it HST} with respect to earth
center was computed by the Flight Dynamics Facility at GSFC with errors less
than 200 meters. The position of the earth with respect to the solar system
barycenter was computed using a routine, SOLSYS, supplied by the U. S. Naval
Observatory (Kaplan et al. 1989). 
All events in the pulsar spectrum (and in the 
background regions) were assigned a (barycentric) arrival time with respect 
to the start of the first exposure. As we found a small drift in the times
recorded in the FITS headers, we used the internal clock from the engineering
mode header to do this.
%
%({\bf JS! Why do we say the engineer clock is better?)}
% because doing the pulse profile analysis with the FITS-header times
% basically smeared the peak of the profile
%
To determine the period we folded the arrival times modulo a grid of
test frequencies, 
$f_{i}$, and corrected for the slowdown rate of the pulsar. 
Pulse profiles were calculated as histograms
of the function $f(t)=f_{i} t + \dot{f} t^{2}/2$, where the data were
coadded into 512 phase bins.
The appropriate value of $f_{i}$ was then determined by 
maximizing the sum of squares of the values in the pulse profile.

In this procedure we used a value 
for $\dot{f}$ from radio observations at Jodrell bank (Lyne, Pritchard, \&
Roberts 1999), while the 
second time derivative is unimportant for this purpose.
This resulted in a measured period of P=33.492675 ms at Modified Julian Date 
(MJD) 51200.549, the time of the beginning of our first observation 
(number O4ZP01010). 
As the data were obtained during a time period of eight hours, 
we could also determine the pulsar slowdown rate from our observations. 
We obtained $\dot P = (4.0\pm0.4) \times 10^{-13}$ s s$^{-1}$,
which is consistent with the value used above from 
the radio observations, $4.2\times 10^{-13}$ s s$^{-1}$.
% NB need 6 decimal accuracy to get this accuracy on Pdot

In Figure 2 we show the Crab pulsar pulse profile in the FUV regime. It was 
obtained by subtracting the background from the pulse profile obtained for 
the period given above. For comparison, the figure also shows the NUV pulse
profile from G98.

\subsubsection{Phase-averaged spectrum}

Averaging over the pulse we obtain a phase-averaged spectrum of the
pulsar which covers the region $1140 - 1720$~\AA. 
This spectrum probes the emission of the Crab pulsar further 
into the UV than has previously been done and is
shown in Figure~3, together with the NUV spectrum of G98.
These spectra 
overlap nicely and cover together the whole range from $1140-3200$~\AA. 
The combined spectrum offers the possibility to deduce the
amount and characteristics of the interstellar reddening, as well as
to determine the dereddened pulsar spectrum itself. 

The FUV-spectrum
was extracted with a 13 pixel wide window. The reductions were made using the
CALSTIS software developed at the GSFC. 
These IDL-routines flatfield the images and then 
the point source spectrum is localized and traced on the detector. 
The extracted pulsar spectrum is background subtracted and converted to 
absolute flux units using the G140L sensitivity table.
The accuracy of the absolute flux calibration is $\sim 15\%$ 
over the full wavelength scale. 
% got the 15% from Don (priv.comm), it is the absolute uncertainty in the 
%white dwarf model, this is the possible offset/difference to optical. the 
%repeatability of G140L is better, a few %, also my countstatictics is
%better (can do a lot of  binning for the fluxlevel) at least everywhere 
%except farfarfarUV
Wavelengths are assigned from a library dispersion solution, while 
the zero point adjustments are determined from arc frames taken 
through the $0\farcs05$ slit for each science 
observation. The wavelengths are then converted to heliocentric 
wavelengths. The accuracy of the wavelength solution is about 0.4 \AA.
% wavelength 0.3  + o.3 as we did no peak up +0.12 from differences to #6
% ->0.44AA
% note for accuracy
% flux, the 6 different exposures gave slightly different fluxes. rms can give 
% a measure of this in the extracted spectrum I have added errors from 
% extraction procedure as well as from RMS of 6 observations
% wavelengths, I have simply used wavelenghtsscale for observation #6 !
% this is similar to crude binning, must be accounted for in errorestimate 
% above

\subsection{Optical observations}
In addition to the UV spectrum, we have also collected data in the
optical regime. During several nights in December 1998 we did spectroscopy
of the Crab pulsar using the Andalucia Focal Reducer and Spectrograph (ALFOSC)
at the 2.56m {\it NOT} on La Palma. 
In total we obtained 11.25 hours of data in five 
different grisms.
The 1\farcs2 slit was used for all observations (see Table 2 for more details).
Not all nights were photometric and the seeing was generally
just above 1\arcsec. The data were bias subtracted and flatfielded. 
Wavelength calibrations were done using arc frames obtained
with a helium lamp. Flux calibration
of the spectra was accomplished by comparison to
the spectrophotometric standard stars Feige~34 and G191-B2B. To avoid
systematic errors due to background subtraction the slit was put at two
different position angles.
%PA 75 and -70?
All observations were made at low
air masses and close to the parallactic angle to reduce the effects of
atmospheric
dispersion; the Crab pulsar passes just $7\arcdeg$ from zenith as
viewed from La Palma in December.
The slit positions for the {\it NOT} observations are shown in Figure~1. 

\subsection{Combined UV/optical spectrum}

The combined optical and UV observations (both NUV and FUV) are shown in 
Figure~4. It covers the region $1140 - 9250$~\AA. 
Although great care was given to the background subtraction of the 
nebula, the optical spectrum of the Crab pulsar was contaminated by over- 
and under-subtractions of strong nebular emission lines.
% as well as telluric features. 
These had to be taken out by hand, and
we used the IRAF task SPLOT to interactively clean the spectrum. Points that 
deviated more than 4$\sigma$ from a smooth continuum fit were rejected. 
This procedure was robust enough 
to exclude only clear cases of nebular contamination.
The optical spectrum used is a combination of all the different
spectra in Table~2.
Note that an absolute flux calibration
was not applied to the optical spectrum, as 
the observing conditions were often non-photometric. Instead we have applied 
a grey shift to the spectrum to match the $V$-band observations of Percival
et al. (1993) and Nasuti et al. (1996).
As seen in Figure 4, this is well matched with the
UV spectrum from the {\it HST}.

%--------------------%

\section{Discussion}

\subsection{Pulse profile}
The work of Percival et al. (1993) showed small differences in the 
optical versus NUV pulse profile shapes.
They observed the Crab pulsar with the High Speed Photometer (HSP)
onboard {\it HST} and found that the main pulse is slightly narrower in 
the UV than in the optical. 
Eikenberry et al. (1997) extended the analysis into the 
near-IR and found that the trend for a decreasing Full Width Half Maximum 
(FWHM) with decreasing wavelength seems to hold over the full UV-IR range. 

The time-tag mode of STIS/FUV-MAMA allows us to examine if the pulse profile
of the emission is different in this wavelength region from that in the NUV. 
The most striking feature of the FUV pulse profile shown in Figure 2 is
certainly that it is very similar to the profile previously obtained in 
the NUV (G98). It appears, however, that the primary peak is slightly narrower
in the FUV than in the NUV, as indicated
by the blow-up of that region in Figure 2.
We measured the FWHM of the 
FUV and NUV primary peak to be 0.0405 and 0.0426 periods, respectively.
The position of the peak was determined by a polynomial fit to the central 
10 phase bins. The phase at half-maximum was then simply determined by linear 
interpolation between the two closest phase bins. 
%We note that our 125 $\mu$s
%resolution does not resolve the peaks, which might introduce some smearing
%of the peaks. As the NUV and FUV data were obtained 
%and reduced in exactly the same way, 
%we do, however, believe that the difference is significant.
To estimate a statistical
error on the procedure used to measure the FWHM, we computed
the FWHM for each of our six FUV observations. 
The standard deviation obtained in
this procedure was 0.001 phase bins. The measured difference in the FWHM of
the primary peak can therefore be considered marginally significant.
% sqrt[(mean-fwhm(i))^2/5]/sqrt(5)
The secondary peak in Figure~2 might actually appear broader in the FUV.
It is, however, much noisier than the primary peak and
the same 
procedure as above could not determine any significant difference to the
13$\%$ (3$\sigma$) level.
The above findings are in agreement with the trend seen in Percival et al. 
(1993) and Eikenberry et al. (1997).
The pulse period obtained from the FUV observations is P=33.492675 ms.
Radio data from Jodrell Bank (\cite{LPR99}) determined the pulse period 
to be P=33.492402 ms on January 15 1999. Using their values for the pulse 
period and its first time derivative on this date we can calculate the 
period at the time of our {\it HST} observations. 
The result is P=33.492676 ms.
%http://www.jb.man.ac.uk/~pulsar/crab.html
This agrees with our estimate to 7 significant digits. 
The limiting errors in our computation of the
period are the accuracy of the SOLSYS routine which has a barycentric velocity
error of less than $8.0\EE{-7}$ AU/day and the unknown accuracy of the rate
of the STIS onboard clock.

\subsection{The UV extinction curve}

The phase averaged UV spectrum of the Crab pulsar must be corrected
for a substantial amount of interstellar reddening.
The value for $E(B-V)$ has been estimated by a number of authors; 
Wu (1981) obtained $E(B-V)=0.50\pm0.03$ mag by using the 2200 \AA\ dust
absorption feature, the nebular synchrotron continuum and an extinction
curve derived from eight stars. Blair et al. (1992) also used the best
fit to the UV nebular continuum and obtained
$E(B-V)=0.51^{+0.04}_{-0.03}$ mag.
A different method was used by Miller (1973), who determined the
reddening of the Crab nebula from observations of [S II] lines. 
Using modern values for the atomic parameters (Keenan et al. 1993; 
Ramsbottom, Bell, \& Stafford 1996), and the extinction curve of 
Fitzpatrick (1999), his measurements gives
$E(B-V)=0.50^{+0.04}_{-0.06}$ mag.

Our data of the pulsar itself allows us to estimate the value for $E(B-V)$
by ``ironing out'' the 2200 \AA\ bump.
To do this we assumed a standard value $R=3.1$ and
dereddened the UV spectrum for different values of $E(B-V)$. 
The dereddened spectra were then fitted by a power law and we chose 
the value for $E(B-V)$ that minimized $\sigma_{\alpha_{\nu}}$,
the standard deviation of the power law fit in the region 
log $\nu$=[14.98, 15.41],
where the region including the Ly$\alpha$ absorption was excluded.
This procedure used the
galactic mean extinction curve from Fitzpatrick (1999) and gave
$E(B-V)=0.52$ mag.
This is in excellent agreement with the previous results stated
above. As our data have better signal-to-noise and sampling than
previous continuum fits, we 
will use this value as the best estimate of the extinction toward
the Crab nebula throughout this paper. 

% wu gets EBV=0.50 +- 0.03
% miller 73 says Av=1.6+-0.2 which gets EBV=0.516+-0.07 for R=3.1
% blair says ebv=0.51 +0.04 -0.03 from UV continuum fit
%; davidson fesen says: (based on miller and Wu)
%; av=1.46+-0.12 and ebv=0.47+-0.04  
%;this would mean Rv=3.106
%; 1.46+0.12 / 0.47-0.04 = 3.6744 ; 1.46-0.12 / 0.47+0.04 = 2.6275
%; thats a bit pessimistic 3.11+-0.13 is rootmeansquareerror

From the NUV spectrum
taken of the Crab pulsar by {\it IUE}, claims were made for a peculiar
extinction curve (\cite{Ben80}). In particular, the 2200 \AA\ bump
was reported to be substantially narrower than for the galactic mean 
extinction curve,
a finding that could indicate that the
supernova event itself had altered the grain composition in the
Crab nebula.
In principle both the extinction curve and the
intrinsic pulsar spectrum are unknown, which of course makes it
troublesome to disentangle these quantities.  We take the following
approach to this problem: theoretical models favor a power 
law spectrum (e.g., \cite{LGS98}) and dereddening with 
$E(B-V)=0.52$, $R=3.1$ indeed gives a power law pulsar spectrum (Fig. 4), 
so we will simply assume the intrinsic spectrum of the Crab pulsar to
follow a power law $F_{\nu}$ = $K(\nu/\nu_{0}$)$^{\alpha_{\nu}}$
ergs~s$^{-1}$~cm$^{-2}$~Hz$^{-1}$.  Here $\nu$ is the frequency of the
radiation and $K$ is a constant that is nearly independent of
$\alpha_{\nu}$ when $\nu_{0}$ 
is the logarithmic mean frequency of the fitted bandpass (\cite{P93}).  
Using the extinction parameters above, $E(B-V)=0.52$ and
$R=3.1$, the spectral index is $\alpha_{\nu}=-0.035$ in the UV range.
%To find the extinction
%parameters E(B-V) and R we dereddened the UV  spectrum for a fine grid
%of values in the range E(B-V)=[0.4,0.7],  R=[2.5,5.0], and chose the
%parameters that minimized $\sigma_{\alpha_{\nu}}$,  the standard
%deviation of the power law fit in the region log $\nu$=[14.98,15.41],
%where the region including the Ly$\alpha$ absorption was excluded.
%At this stage, we have used only the UV data, as these are more
%sensitive to the extinction and less affected by atmospheric
%dispersion and nebular contribution.  This procedure, that used the
%galactic mean extinction curve from Fitzpatrick (1999), gave E(B-V)=0.514
%and R=3.70 for a power law with  spectral index
%$\alpha_{\nu}$=$-0.66$. Although E(B-V) seems to be well constrained by
%this procedure, the value of R is not. This hampers the determination
%of  $\alpha_{\nu}$, see the discussion in section 3.3.  

By assuming that the intrinsic pulsar spectrum is indeed well represented by 
the obtained power law we derived
an extinction curve toward the pulsar. This is shown in
Figure 7 together with the mean galactic extinction curve for $R=3.1$ from
Fitzpatrick (1999). The derived extinction curve has a moderately narrower dip
(15$\%$) 
and a somewhat shallower rise in the extreme FUV. Apart from this, it
overlaps nicely with the galactic mean extinction curve.
% for the same value of $R$. 
Considering the large variety of measured UV
extinction curves (see Fig. 2 in Fitzpatrick 1999), 
the derived extinction curve
toward the Crab can hardly be claimed to be peculiar. 
Assuming that the pulsar spectrum follows a power law, we conclude that
we find no evidence for a non-standard extinction curve toward the
Crab pulsar.

\subsection{The spectral index of the pulsar continuum}

According to models, the high energy pulsar emission, from IR to
$\gamma$-rays, is produced by (curvature) synchrotron radiation 
(e.g., \cite{LGS98}, and references therein). 
A number distribution of 
electrons following a power law
$N(E) dE = C E^{-\gamma} dE$, where $E$ is the energy, $C$ is a constant and 
$\gamma$ is the electron spectral index, will produce synchrotron 
radiation that has a power-law distribution in flux density
$F_{\nu} = K(\nu/\nu_{0})^{\alpha_{\nu}}$ 
ergs~s$^{-1}$~cm$^{-2}$~Hz$^{-1}$.
%(Shklovsky 1960).
The photon spectral index, $\alpha_{\nu}$, is related to the electron
spectral  index, $\gamma$, via $\alpha_{\nu} = -(\gamma-1)/2$ for
synchrotron radiation. 
% took this from hill et al., the def. of alpha is standard and we
% must stick to it for comparison with oke, percival etc.
% def of -gamma also pretty standard, I guess,
% NB I think hill et al. makes a mistake in their formula relating
% alpha and gamma, I think the above is correct (JS)

The early low resolution spectroscopy of Oke (1969) appears to peak in the
middle of the observed region ($\sim 3400-8000$ \AA). He reports a slope of 
$\alpha_{\nu} = -0.2$ with no stated errors, although he cautions that the 
uncertainty in the reddening correction could allow also for a positive slope. 
Much of the theoretical work on the optical emission mechanism of pulsars 
has been based on this finding 
(cf. Ginzburg \& Zheleznyakov 1975; \cite{LGS98}).
More recently, Percival et al. (1993) used 
ground-based optical broadband photometry together with a 
NUV photometric point 
from {\it HST}/HSP to determine a slope of $\alpha_{\nu}=0.11\pm0.13$, while
Nasuti et al. (1996) obtained a spectrum in the limited wavelength 
range $4900-7000$ \AA\ (see below) and 
determined $\alpha_{\nu} = -0.10\pm0.01$.
%While the minimization procedure above gave a reasonable value 
%for E(B-V) the obtained value for R appears rather high. As mentioned, this 
%parameter is in fact not very well constrained in this procedure. Allowing
%the RMS of the power law fit to increase by 1$\%$ above the minimum gives the
%allowed parameter surface E(B-V)=[0.50,0.53], R=[3.3,4.2]. 
%If we would force R to the standard value R=3.1, we obtain E(B-V)=0.519
%and a power law fit to the UV region for these parameters gives 
%$\alpha_{\nu}$=-0.038.

Applying $R=3.1$ and $E(B-V)=0.52$ from \S 3.2 to
our UV spectrum gives a spectral index of
$\alpha_{\nu}=-0.035\pm0.040$, 
where the error is simply the RMS around the fit. 
Obviously, the main uncertainty in this procedure is the extinction 
parameters. For $E(B-V)$ in the range [0.48, 0.55], 
the uncertainty due to reddening becomes 
$\alpha_{\nu}=-0.035^{+0.094}_{-0.13}$.
An accurate estimate of the spectral index obtained in the UV,
for the given extinction range, can be obtained by 
$\alpha_{\nu}=-0.035+3.12[E(B-V)-0.52]$.
%correcting the spectral index obtained above for
%$-1.24 \delta R + 3.12 \delta E(B-V)$, where $\delta R=R-3.1$ and 
%$\delta E(B-V)=E(B-V)-0.53$. 
%This linear estimate
%, where $\alpha_{\nu}$ becomes more positive for larger 
%$E(B-V)$ and smaller R, 
%is accurate to within 0.001 units of $\alpha_{\nu}$ in
%the above mentioned extinction range.
Note that the spectral 
index for $E(B-V)=0.55$ was erroneously reported in G98.
The analysis in this paper supersedes this previous report.

Including the optical spectrum from the {\it NOT} 
gives a wider wavelength 
range for the fit. 
The dereddened spectrum shown in Figure 4 was obtained with $R=3.1$ and
$E(B-V)=0.52$. The best power law fit to the complete spectrum 
(excluding Ly$\alpha$)
% and the bluest optical part (see Fig. 6)) 
gives $\alpha_{\nu}=0.11$.

Using the full wavelength range we have also tried to constrain 
both $E(B-V)$ and $R$. This can be done using the extinction curves
from Fitzpatrick (1999), which are a one-parameter family in $R$.
By assuming an intrinsic power law we thus allowed both $R$ and $E(B-V)$ to 
vary, and chose the values that minimized $\sigma_{\alpha_{\nu}}$, 
the standard deviation of the power law fit.
This procedure gives $R=3.0$ and $E(B-V)=0.51$,
which is consistent with the values used above. 
In Figure 6 we show the dereddened spectrum of the Crab pulsar for several 
different values of $R$ and $E(B-V)$. In this figure we have also included IR 
data from Eikenberry et al. (1997). These plots clearly show the ambiguity
of the reddening correction. 

We can express the spectral index of the power law fit for the 
values $E(B-V)=0.52$ and $R=3.1$ as $\alpha_{\nu}=0.11\pm0.04^{+0.21}_{-0.22}$.
The first error represents the RMS around the power law fit, 
and the last errors include all power law fits in the extinction 
intervals $R=[2.9, 3.3]$, $E(B-V)=[0.48, 0.55]$.
A linear fit, 
%$\alpha_{\nu}$=0.11-0.38$(R-3.1)$+3.88[$E(B-V)-0.52$]
$\alpha_{\nu}=0.11-0.38(R-3.1)+3.88[E(B-V)-0.52]$
reproduces the obtained $\alpha_{\nu}$ in this interval to better 
than 0.02 units.
% largest error is 0.0122

We can only echo Oke (1969) in his conclusion that the 
uncertainties in reddening corrections are large enough to allow both for
negative and positive slopes.
The inferred energy distribution for the electrons is 
given by $\gamma = 0.8 \pm 0.5$.
It is worth noting that the only other young pulsar for which a spectrum has
been obtained in the optical, PSR B0540-69, had $\alpha_{\nu}=-1.6\pm{0.4}$ 
(\cite{H97}). This is clearly steeper than the Crab pulsar spectrum.

Finally, the time tag mode of our FUV observations also allows us to extract 
phase-resolved spectra, and we have looked for spectral differences during 
the pulse phase. We found no significant differences above the $5 \%$ level, 
neither in the spectrum of the primary versus the secondary peak, 
nor in the leading versus trailing part of the primary peak.

\subsection{The $5900$~\AA\ feature}
In the first report on the optical spectrum of the Crab pulsar,
Minkowski (1942) reported a featureless continuum with no
absorption or emission lines. The observations of Oke (1969) indicated a
similar conclusion, although the spectral resolution was not very good. Since
then, few attempts have been made to obtain a better optical spectrum of 
the Crab
pulsar, despite that technical development certainly admits
improvements. The only modern optical spectrum of the Crab pulsar was 
obtained with the {\it New Technology Telescope (NTT)} by 
Nasuti et al. (1996). 
This phase-averaged spectrum covered a rather small wavelength 
region ($4900-7000$~\AA) and was flat with $\alpha_{\nu} = -0.10\pm0.01$ 
if dereddened with $E(B-V)=0.51$. 
The spectrum was also reported to show a
large dip at $\sim5900$~\AA, the width being $\sim100$ \AA. 
According to the authors this feature probably originates close to the pulsar
itself, but no detailed physical mechanism was proposed.

We have searched for this feature in our optical spectra, but found no
sign for a dip. This is true for all the different gratings covering
this region during all of the nights of data obtained in December 1998 
(see Fig.~4). 
Neither do we see the feature in unpublished
data taken by us at {\it NOT} one year earlier (December 1997). Nasuti et
al. (1996) noted that the feature became enhanced after
flux calibration. This suggests that the dip may be an artefact of the 
reduction procedure. While their investigation used a single 
spectrophotometric 
standard star
with flux sampled only every 100 \AA, we have
used two standard stars with 2 \AA\ sampling in the relevant
wavelength region.
Although the feature could be time dependent, we propose
to regard it as an artefact until confirmed by other observations.

\subsection{Absorption lines}

As in the NUV spectrum of the pulsar (G98), the FUV spectrum contains several 
absorption lines. The identified lines are shown in Figures 7 and 8, and 
their equivalent widths (EWs) and
corresponding column densities, $N$, are listed in Table 3. 
For completeness, we have also included the lines 
in the NUV spectrum identified and measured by G98.

The strongest line is Ly$\alpha$, and its damping wings can be used to 
estimate $N$(H~I). To do this we
assume that the optical depth in the damping wings is defined 
by $\sigma(\lambda)N$(H~I) (Shull \& Van Steenberg, 1985). 
Here the absorption cross 
section $\sigma(\lambda) = (4.26\EE{-20}~{\rm cm}^2)/(\lambda - \lambda_0)^2$,
where the wavelengths are in \AA\, and $\lambda_0 = 1215.67$~\AA. We have
not considered the instrumental profile, which is much narrower ($\sim 1$~\AA)
than the line width (see Fig. 7). From this analysis we
obtain $N$(H~I)$ = (3.0\pm0.5)\EE{21}$ cm$^{-2}$.
The spectrum corrected for Ly$\alpha$ absorption is shown in Figure 7. Our 
estimated uncertainty in $N$(H~I) reflects the uncertainty in the 
continuum fit of the corrected spectrum. The column density of
free electrons toward the Crab is measured from pulsar dispersion to
be  
$N_{\rm e}$~$=~(0.1755\pm0.0007)\EE{21}$~cm$^{-2}$~(\cite{C69}). 
The amount of atomic
hydrogen is thus $N$(H)$ = (3.2\pm0.5)\EE{21}$ cm$^{-2}$.
This value agrees well with 
both the result of Schattenburg \& Canizares (1986) 
who obtained $N$(H) $= (3.45\pm0.42)\EE{21}$ cm$^{-2}$ from an estimate
of the X-ray absorption toward the Crab, and the value we estimate 
from $E(B-V)$ using the relation in de Boer, Jura, \& Shull (1987), 
which gives $N$(H~I)$\simeq 3.0\EE{21}$ cm$^{-2}$ for $E(B-V) = 0.52$. 

The other absorption lines we identify around zero velocity in the FUV 
spectrum are: C~I~$\lambda\lambda$ 1277,1329,1561,1658, C~II~$\lambda$1335,
O~I~$\lambda$1303, Al~II~$\lambda$1671, Si~II~$\lambda\lambda$ 1260,1527 and 
Si~IV~$\lambda$1394. The Si~IV line is close to noise level, which explains 
why the Si~IV~$\lambda$1403 component is not seen. 
The C~I~$\lambda\lambda$ 1277,1329 lines are also marginal detections, but 
their strengths correspond to those expected when compared with the 
strengths of C~I~$\lambda\lambda$ 1561,1658. 
The C~IV~$\lambda\lambda$ 1548,1551 
doublet is absent at zero velocity, but shows a blueshifted absorption with a 
maximum shift of $\sim 2500 \kms$. We return to this line in \S 3.7.

All lines except Ly$\alpha$, and  
the C~IV doublet which is not 
interstellar or from a slow wind (\S 3.7), are unresolved by the 
moderate spectral resolution in 
the STIS spectra ($\approx 1.2$ \AA\ in FUV and $\approx 3.2$ \AA\ in NUV, 
corresponding to $250 \kms$ and $400 \kms$ for the central wavelengths of the 
two gratings, respectively). 
The measured EWs of the absorption lines are rather large, which
means that we
cannot assume that the lines are resolved and optically thin
(i.e., we cannot use 
``weak-line'' theory [e.g., Morton 1991]) to derive abundances of 
the various species, as this would severely underestimate the abundances. 

In the absence of a proper model for the distribution of intervening matter, 
we assume that the absorption is dominated by a single cloud component. We 
set the intrinsic ``Doppler'' width in this component to $1 \kms$, and assume 
that this is the same for all lines.
% (which, e.g., would be the case for turbulence). 
Furthermore, we consider only one spectral component of the lines listed in 
Table~3. All these assumptions cause us to systematically overestimate the 
column densities and abundances so that our estimates will be upper
limits to these. With this in mind, we calculate 
Voigt line profiles and EWs as functions of column density, using the atomic 
data in Morton (1991). From the measured EWs we then obtain column densities 
and abundances for the interstellar gas toward the Crab. The abundances in 
Table~3 are presented in the standard logarithmic form where the value for 
hydrogen is set to 12.0. 

In Table~4, where we have simply coadded the abundances of the different 
ionization stages of each element, we give the abundances of C, O, Mg, Al, Si 
and Fe from our analysis. We also list the solar values for these elements
according to Anders \& Grevesse (1989). 
There appears to be no extreme 
depletion of any element, contrary to what was stated in G98, 
where ``weak-line'' theory was used. 
This is in agreement with the X-ray observations 
by Schattenburg \& Canizares (1986) which were consistent with a solar
abundance of oxygen.
Our oxygen abundance from the single O~I~$\lambda$1303 line is, however,
rather uncertain due to geocoronal airglow corrections, 
and a possible blending with 
Si~II~$\lambda$1304.
As pointed out above, our method is likely to systematically 
overestimate abundances. Broader lines, or more cloud components,
%than assumed above 
would lower the abundances derived in accordance with 
depletion seen in the normal interstellar medium (e.g., 0.65 dex for carbon
at 2  kpc, Welty et al. 1999).
Spectra with higher resolution are required to refine this analysis.

\subsection{Emission lines}
%Although our study concentrates on the pulsar itself, t
The FUV-spectrum contains
also information about the Crab nebula. The long slit 
covers 
$\sim$ 25\arcsec $\times$ 0\farcs5 of the nebula, along position 
angle (PA) $40\arcdeg.5$.
% Pa aper is the correct one to use
%PA-APER =   0.405058677856E+02  position ang of aperture used with target(deg)
%ORIENTAT=              40.8892  position angle of image y axis (deg. e of n)  
The orientation of the slit is shown in Figure 1.
To extract the pulsar spectrum 
we used only 13 of the 1024 pixels of the MAMA detector. 
To obtain a spectrum of the nebular emission outside this extraction window 
we again used CALSTIS to produce rectified, wavelength-  and flux-calibrated 
images. As the count rates were very low, we summed the
emission from two 8\farcs2 long regions positioned 0\farcs66
above and below the pulsar position. This safely excludes any contribution 
from the point spread function (PSF) wings of the pulsar. 
All emission from all 6 observations was combined and 
averaged to increase the 
signal-to-noise. We present this nebular spectrum in 
Figure 9, which shows only the wavelength region $1400 - 1700$ \AA. 
This is to exclude the %strong 
geocoronal lines of Ly$\alpha$ and 
O~I~$\lambda\lambda$ 1303,1356. %{\bf PL: do we see the 1356 component?} 
The spectrum is the average of the regions above and below the pulsar.
For the spectrum extracted above (North-East of) the pulsar, the
continuum level was flattened artificially. This is because
we observed a rising continuum in the red part of this spectrum,  
which is probably due to contamination
of the nearby star seen in Figure 1.
Moreover, 
the count rates in the continuum are only $\sim 7.5 \times 10^{-5}$ 
counts pixel$^{-1}$ s$^{-1}$.
At such low count rates, trends in the dark currents might influence the 
continuum. No dark images are subtracted 
from the FUV-MAMA observations, because the 
dark images are known to be variable
and to have low count statistics.
Therefore, we will not discuss further the slope of the continuum emission
from the nebula. 

Only two emission lines intrinsic to the Crab nebula can be
seen, C~IV~$\lambda$1550 and He~II~$\lambda$1640.
The measured intensities
are $\sim (2.1\pm0.8)\times10^{-15}$ and $\sim (1.4\pm0.8)\times10^{-15}$ 
ergs~s$^{-1}$~cm$^{-2}$, respectively. This is an average for the two
8\farcs2 $\times$ 0\farcs5 regions, just above and below the pulsar. 
Our detection of these two lines
is in accordance with the two previous spectroscopic 
observations of the Crab nebula in the FUV. 
Davidson et al. (1982) used the {\it IUE} to detect 
these two lines, as well as C~III]~$\lambda1909$. 
Blair et al. (1992) recalibrated the {\it IUE} data and complemented
these with 
observations from the {\it Hopkins Ultraviolet Telescope (HUT)}. 

The widths of the lines are approximately 13~\AA~and 10~\AA,
for C~IV~$\lambda$1550 and He~II~$\lambda$1640, respectively.
The width is due to the velocity distribution of the material, the
doublet nature of C~IV~$\lambda$1550 as well as to the spectral
resolution for extended sources for the spectrograph.
The bulk of the emission is redshifted by $\sim 1000~\kms$,
although a narrower zero-velocity component is seen from the region
above the pulsar.

The flux ratio of C~IV~$\lambda$1550 to He~II~$\lambda$1640
is thus $\sim1.5$, which is in accordance with the  
findings of Davidson et al. (1982), who sampled the fluxes over 
a larger field, $20\arcsec \times 10\arcsec$. They used this
to argue that the Crab nebula has no overabundance of carbon. However, 
the subsequent investigation by Blair et al. (1992) showed that variations in 
the He II/C IV line ratio exist in the nebula. They cautioned on conclusions
regarding the carbon abundance, as differences in physical parameters (e.g.,
ionization, density, temperature, clumping) are difficult to disentangle from 
abundance variations. 
%In fact, variations on these line ratios can almost 
%be seen directly on our 2-dimensional frames. 
To determine the carbon abundance is important because it holds the potential
to provide information about the 
ZAMS mass of the progenitor. 
Comparisons to detailed photoionization 
models for the large apertures used by {\it IUE} and {\it HUT} are hampered by 
the fact that ionization conditions are known to change over very small spatial
scales (\cite{Sa98}). The observations we present have finer spatial resolution
and are not biased toward bright filaments. {\it HST} 
observations {\it of the same regions} in the optical, to establish the 
ionization conditions, could make more quantitative estimates of the 
carbon abundance possible.

\subsection{The outer shell}

As was mentioned in \S 3.5, and shown in Figure 8, C~IV~$\lambda$1550 is the 
only line in the UV spectrum which shows clear evidence of blueshifted 
absorption. No evidence for absorption can be seen at zero velocity. The 
greatest absorption appears to arise in material moving at $\sim 1200 \kms$, 
and can thus 
be due to ``normal'' Crab nebula material. However, the absorption
seems to continue to higher velocities, and can be traced out to 
$\sim 2500 \kms$ (see Fig.~10). 
To estimate the significance of the detection we have calculated an average
and RMS from two 50~\AA~regions redward and blueward of the C~IV
line. Although the line is too noisy for individual pixels to be
significant, it consists of many pixels that are consistently below
the average. The eleven pixels between $1650 - 2780~\kms$
give a detected absorption with a significance of $5.8\sigma$.
For the five points in the region $2330 - 2780~\kms$
the significance is $3.0\sigma$.

Together with Clark et al. (1983), who also reported velocities
in excess of $2000 \kms$, this is the highest velocity ever measured 
in the Crab, and can be interpreted as evidence for the existence of the long 
sought fast outer shell (cf. \S 1). The column density of C~IV in 
Table 4, $N($C~IV$) = (3.0\pm1.1)\EE{14}$ cm$^{-2}$, 
assumes the line to be resolved and optically thin. 
This should be the case for velocity broadening caused by the tentative fast 
shell, though we caution that the material moving at $\sim 1200 \kms$ may 
not be spectrally resolved. The total column density of C~IV is therefore 
likely to be higher, while that of the proposed fast shell is lower
than the value given in Table 4. We will now investigate whether the detection 
is consistent with a fast outer shell model and what information can be 
provided about the supernova ejecta.

\subsubsection{Constraints from C~IV~$\lambda$1550}

We adopt a model similar to that of LFC86, i.e., outside the 
observed filaments we attach a massive, 
freely coasting, spherically symmetric shell. 
The inner radius, $R_{\rm in}$, is set to $5.0\EE{18}$ cm, which agrees
with the ``mean'' inner radius of the presumed shell used by Sankrit \& 
Hester (1997, see their Fig.~7). For free expansion this corresponds to a 
velocity of $\approx 1680 \kms$. Furthermore, we 
assume that the mass of the shell, $M_{\rm sh}$, is $4 \Msun$, and that the 
outer radius of the shell, $R_{\rm out}$, is $1.9\EE{19}$ cm. The maximum 
velocity, $V_{\rm out}$, is then $\approx 6370 \kms$, and the 
total kinetic energy of the shell, $E_{\rm sh}$, is $1.0\EE{51}$ ergs, if the 
density is constant in the shell. For a density which decreases with 
radius, $E_{\rm sh}$ is lower, if $M_{\rm sh}$ is held constant. This is
shown in Table 5 for density slopes up to $\eta = 9$, where $\eta$ is defined
as $\rho(R) = \rho(R_{\rm in})~(R/R_{\rm in})^{-\eta}$.

In our model we have used the relative 
abundances $X($H$)$: $X($He$)$: $X($C$)$ = 1.0: 0.1: $3.5\EE{-4}$, where 
the $X($C$)$/$X($H$)$ ratio corresponds to the solar value.
With these abundances we have calculated the absorption in 
C~IV~$\lambda\lambda$ 1548,1551 as a function of $\eta$
and the relative fraction of carbon in C~IV, $X($C~IV$)$.
The parameter $X($C~IV$)$ is unity when carbon is all in C~IV. 

In Figure 10 we show results for two models where $X($C~IV$)$ has been kept
constant throughout the shell. The dotted line shows the C~IV line in the 
case of $\eta = 0$ and $X($C~IV$) = 1$, while the dashed line is 
for $\eta = 3$ and $X($C~IV$) = 0.14$. Both models are described in Table 5. 
The photoionization models of 
LFC86 show a case similar to our $\eta = 0$ case (their model 1). There,
carbon is ionized beyond C~IV, and only a very small fraction, concentrated 
to the outer edge of the shell, remains in C~IV. This gives maximum absorption 
in the $\eta = 0$ case around $-6000 \kms$ (see Fig. 1 of LFC86 for such a 
case). This is clearly not what is seen in the FUV data. On the contrary, the 
observed absorption peaks at lower velocities, meaning that the C~IV number 
density %, $n({\rm C~IV})$,
must be the highest closer to $R_{\rm in}$. 
The photoionization models of LFC86 make the $\eta = 0$ case rather 
unlikely from this point of view. Instead we turn our attention to 
steeper density profiles.

Such a situation is highlighted by the $\eta = 3$ case in Figure 10. 
Our assumption of constant $X($C~IV$)$ in the shell is probably more realistic 
for $\eta = 3$ than for $\eta = 0$. This is because the ionization 
parameter, $\xi = {\rm n}_{\gamma} / {\rm n}_{\rm e}$, 
where ${\rm n}_{\gamma}$ and ${\rm n}_{\rm e}$ are number densities of
ionizing photons and electrons, respectively, has the radial
dependence $\xi \propto R^{\eta-2}$, if absorption can be neglected. With 
absorption included, $\eta$ must be $>2$ to obtain a near-constant $\xi (R)$. 
Our choice of $X($C~IV$) = 0.14$ has only been made to fit
the data. Any combination of $X($C~IV$)$ and $M_{\rm sh}$, so 
that $X($C~IV$) M_{\rm sh} \approx 0.56 \Msun$, would fit the data equally 
well. In our model, for $\eta = 3$, this gives a minimum mass 
(for $X($C~IV$)=1$)
of the shell of $M_{\rm sh} \sim 0.6 \Msun$, and a minimum kinetic energy 
of $E_{\rm sh} \sim 8\EE{49}$ ergs. 

While a constant $X($C~IV$)$ in the shell gives a good fit to the line
profile for $\eta = 3$, cases with a steeper density dependence must have an 
ionization structure where $X($C~IV$)$ increases outward through the shell.
From the radial dependence on the ionization parameter this could be possible,
but may require a fine-tuning between the $R^{\eta-2}$ and exp$(-\tau)$ parts
of $\xi$. Time dependent effects are also likely to be important, and the
likelihood of an increasing $X($C~IV$)$ with radius can only be explored by
numerical models. We postpone this to a future study. To give the same 
absorption at $2500 \kms$ as the $\eta = 3$ model in Figure 10, models 
with other $\eta$ must have $X($C~IV$) \sim 1.49^{\eta-3}~X($C~IV$)_{\rm in}$. 
The parameter $X($C~IV$)_{\rm in}$ is the value of $X($C~IV$)$ at $R_{\rm in}$,
and is the value given in Table 5.
For example, for $\eta = 9$, $X($C~IV$)$ must be $\sim 11$ times higher at 
the radius of $2500 \kms$ compared to $X($C~IV$)_{\rm in}$.

It thus appears as if the observed line profile of the C~IV~$\lambda1550$ 
absorption is consistent with the outer shell being fast and massive. 
That is, the C~IV line cannot exclude that the shell could carry an energy
of $10^{51}$ ergs, because the absorption in the gas with the highest 
velocities ($\sim 6000 \kms $) in the outer shell model would simply disappear 
in the noise of our rather poor signal-to-noise spectrum.

To limit ourselves to the observed velocities, 
we adopt a maximum velocity of $3000 \kms$ to obtain
lower limits on the mass and energy of the shell.
Table 5 shows that $M_{\rm sh}$
could be very low (i.e., $M_{\rm sh} = M_{\rm trunc}~X($C~IV$)$, 
with $M_{\rm trunc}$ defined in Table 5) if $\eta$ is large and if we only 
use the absorption at $R_{\rm in}$ as a criterion. However, if we also 
require that the absorption at $2500 \kms$ should be the same for 
any $\eta$ as in the $\eta = 3$ model shown in Figure 10, 
the minimum mass for models with $\eta \geq 3$ is given 
by $M_{\rm sh} \gtrsim 1.49^{\eta-3}~ X($C~IV$)_{\rm in}~M_{\rm trunc} \Msun$,
which for $\eta = 3$ (4, 5, 7, 9) 
becomes $M_{\rm sh} \gtrsim 0.27~(0.31,~0.37,~0.55,~0.90) \Msun$.
The kinetic energy corresponding to 
this $M_{\rm sh}$ (i.e., for $V_{\rm out} = 3000 \kms$)
is $E_{\rm sh} \gtrsim 1.5~(1.7,~1.8,~2.4,~3.4) \EE{49}$~ergs.
The limits on $M_{\rm sh}$ and $E_{\rm sh}$ for shallower density profiles
are fixed by the product $X($C~IV$) M_{\rm trunc}$ (cf. Table 5), and 
are $M_{\rm sh} \gtrsim 0.48~(0.36) \Msun$ 
and $E_{\rm sh} \gtrsim 3.1~(2.2) \EE{49}$~ergs, respectively,
for $\eta = 1~(2)$. All these limits scale inversely with the overall 
abundance of carbon.

To summarize the constraints from C~IV~$\lambda$1550, we first emphasize
that the line shows that an outer shell with maximum velocity 
of $V_{\rm out} \sim 2500 \kms$ appears to be present. 
We have a $\gtrsim5\sigma$ detection of velocities between $1650 - 2780 \kms$.
The kinetic energy of the shell depends on its
mass, density structure and extent. 
For a relatively shallow density 
profile ($\rho \propto R^{-3}$, or shallower) 
the kinetic energy of the shell could be as high as $10^{51}$ ergs, 
if the shell is spherically symmetric and extends to velocities higher 
than those we can detect with the obtained signal-to-noise. 
For a maximum velocity of $\sim 6000 \kms$, the mass of the shell required to
obtain this kinetic energy would be $4-8 \Msun$.
A completely flat density profile, however, seems unlikely from the results
of LFC86, both in the fast case (LFC86, model 1), and for a model
similar to our $V_{\rm out} \approx 3000 \kms$ case (LFC86, model~4).

For a density profile steeper than $\rho \propto R^{-3}$ the kinetic 
energy is most likely $< 10^{51}$ ergs since the shell mass should be lower 
than $\sim 8 \Msun$ to be compatible with progenitor models. The lowest 
mass and energy we estimate for a shell with $V_{\rm out} = 3000 \kms$ 
is for $\eta \sim 3$, and are $\sim 0.3 \Msun$
and $\sim 1.5\EE{49}$~ergs, respectively. These values are approximate
as they depend on spherical symmetry, the value of $R_{\rm in}$, and a model 
fit to the line profile of C~IV~$\lambda1548$ at $-2500 \kms$ where the line 
profile is rather uncertain. To distinguish between models with different 
density slopes (see Table 5), photoionization calculations are needed.

\subsubsection{Other constraints on the shell}

There are, unfortunately, no other lines detected in the UV data that can 
constrain our current analysis further. 
Two potentially useful doublets are 
Si~IV~$\lambda\lambda$ 1394,1403 and N~V~$\lambda\lambda$ 1239,1243,
but we cannot identify absorption at high velocities in any of these two 
doublets. The absence of the two doublets is, however, not surprising. 
Silicon should be more highly ionized than carbon, 
and LFC86 found that C~IV~$\lambda1550$ 
should produce significantly stronger absorption than N~V~$\lambda1240$. The 
spectral region around N~V~$\lambda1240$ is also rather noisy and the line 
sits in the damping wing of Ly$\alpha$. Although the absence of the lines 
cannot constrain models in this simple analysis, it can be used
in conjunction with photoionization calculations to test different models.

The tentative outer shell has been 
previously searched for also in the optical. Searches
in [O~III] have been negative for the region outside the observed [O~III]
skin (cf. Fesen et al. 1997, and references therein). This is not surprising
from the point of view of the models of LFC86, where oxygen is more
highly ionized than O~III.

A highly ionized massive shell is bound to give rise to H$\alpha$ emission.
The deepest search for such emission was done by Fesen et al. (1997) who 
found a surface brightness limit in H$\alpha$ 
of $1.5\EE{-7}$~ergs~cm$^{-2}$~s$^{-1}$~sr$^{-1}$. With the dereddening
suggested by Fesen et al. (1997), i.e., $A_{{\rm H}\alpha} = 2.536E(B-V)$, 
and $E(B-V) = 0.52$ (cf. above), the dereddened surface brightness limit 
becomes $5.1\EE{-7}$~ergs~cm$^{-2}$~s$^{-1}$~sr$^{-1}$. We have calculated the
surface brightness in our models to see how it compares with the observed 
limit. We use a temperature of the tentative shell of $2\EE4$~K, 
which is nearly three times higher than that used by Fesen et al. (1997) and 
Murdin (1994), but in accordance with the models of LFC86. 

The maximum (dereddened) surface brightness in
H$\alpha$, $\Sigma_{{\rm H}\alpha}$, occurs at the impact 
parameter, $p = R_{\rm in}$, i.e., just at the edge of the observed nebula. 
The value of $\Sigma_{{\rm H}\alpha}$ for this impact parameter is given as
a function of $\eta$ in Table 5 for the model with $M_{\rm sh} = 4 \Msun$.
It is seen that the modeled surface brightness exceeds the observed
limit for $\eta \gtrsim 4$. However, $\Sigma_{{\rm H}\alpha}$ decreases
rapidly with $p$ for large $\eta$. Table 5 shows that 
at $p = 1.1 \times R_{\rm in}$ (corresponding to $\sim 17\arcsec$ 
outside the observed nebula for a distance of 2 kpc, and roughly
where the search by Fesen et al. (1997) was conducted), 
the surface brightness exceeds the observed limit only 
for $\eta \gtrsim 5$. For a shell mass as low as $0.3-0.9 \Msun$, which we 
found to be likely lower limits to the shell mass 
for the density slopes investigated, the shell 
would easily have escaped detection in H$\alpha$.
% with the current observational limit. 

A method to derive parameters for the outer shell was devised by
Sankrit \& Hester (1997). They estimated the density needed to form a
radiative shock at the interface between the nebula and the presumed outer
shell, as such a shock is needed in their model to explain the observed
[O~III] skin. They estimate that a minimum density 
of $\rho/{\rm m}_{\rm H} \sim 12$~cm$^{-3}$ is needed, at least in the 
presumed equatorial plane of the nebula. If this is true also in the 
direction toward the pulsar, the models of Sankrit \& Hester 
yield $N($C~IV$) \sim 10^{14}$~cm$^{-2}$
for the radiative tail of the shock. 
This would escape detection in our data since the intrinsic line width in 
their model should be small (much less than our spectral resolution). We can 
therefore not distinguish between a radiative or adiabatic shock (or no
shock at all) in the direction to the pulsar. This also means that it is 
unlikely that any of the absorption we detect occurs in a region similar to 
the radiative region in the model of Sankrit \& Hester.

Sankrit \& Hester (1997) estimate that a mean density at the inner edge
of the shell, averaged over all polar angles, should be $\sim 8$~cm$^{-3}$. 
This would correspond to $n_{\rm H}(R_{\rm in}) \approx 5.7 \cm3$ for 
the He/H ratio used in Table 5. 
To see if we can make a consistent model including the C~IV line,
the H$\alpha$ surface brightness limit by Fesen et al. (1997) and the model 
by Sankrit \& Hester (1997), we have assumed an upper limit to the shell 
mass of $8 \Msun$ (inside $V = 3000 \kms$), 
and used the information in Table 5. We then find 
that $n_{\rm H}(R_{\rm in}) \gtrsim 5.7~(M_{\rm trunc}/8 \Msun) \cm3$
is required 
to get a high enough density at $R_{\rm in}$ to agree with the model of
Sankrit \& Hester (1997). According to the values in Table 5 this is 
fulfilled for $\eta \gtrsim 3$. The mass and kinetic energy for a shell with 
such a high $n_{\rm H}(R_{\rm in})$ and with $V_{\rm out} = 3000 \kms$ would 
be $6.8~(5.0, 3.8, 2.4, 1.7) \Msun$ 
and $3.9~(2.7, 1.8, 1.0, 0.7)\EE{50}$~ergs, respectively,
for $\eta = 3$ (4, 5, 7, 9). A caveat for this model is 
that $\Sigma_{{\rm H}\alpha}$ at $p = 1.1 \times R_{\rm in}$ then
becomes $2.0~(1.4, 1.0, 0.60, 0.35)\EE{-6}$~ergs~cm$^{-2}$~s$^{-1}$~sr$^{-1}$
for $\eta = 3$ (4, 5, 7, 9), which is close to, or higher than, the observed
limit. In this scenario, a limit on the H$\alpha$ surface brightness improved
by a factor of a few, close to the observed nebula, 
should be able to distinguish between models with 
different density slopes even if the He/H ratio is higher than we have
assumed. 
%In particular, if the search for H$\alpha$ emission is made very 
%close to the observed nebula, and close to what may be the equatorial 
%plane. 
Photoionization models to accurately calculate the temperature 
and to check the radial dependence on $X($C~IV$)$, are also needed.

\section{Conclusions}

Using STIS onboard the {\it HST} we have observed the 
Crab nebula and its pulsar 
in the far-UV ($1140-1720$ \AA). We have obtained the pulse profile of 
the pulsar, which is very similar to our previous near-UV profile, although
the primary peak appears to be marginally narrower than in
the near-UV data ($5\%,2\sigma$). 
Combining the far- and near-UV data, and assuming 
an intrinsic power law for the pulsar continuum, we have derived an 
extinction of $E(B-V)=0.52$ mag toward the Crab. No evidence for a non-standard
extinction curve was found. We have also added optical spectra taken with 
the {\it NOT} to obtain a spectrum of the pulsar from 1140 \AA~to 9250 \AA. 
We have shown that the pulsar spectrum can be well fitted over the full 
UV/optical range by a power law with spectral index $\alpha_{\nu}=0.11$. 
The exact value of the spectral index is, however, sensitive to the 
amount and characteristics of the interstellar reddening, and 
we have investigated this dependence for a likely range of $E(B-V)$ and $R$.
In the optical,
we find no evidence for the dip in the pulsar spectrum around 5900 \AA\
reported by Nasuti et al. (1996).

The interstellar absorption lines detected in the UV have been
analyzed, and are consistent with normal interstellar abundances. The
column density of neutral hydrogen is $(3.0\pm0.5)\EE{21}$~cm$^{-2}$, which
corresponds well to the value derived for $E(B-V)$. 
From the Crab nebula itself we detect the emission lines 
C~IV~$\lambda$1550 and He~II~$\lambda$1640. The ratio of the fluxes of these
lines is similar to what has been derived previously, although obtained 
with much improved spatial resolution.
%our field of view was considerable smaller than in the earlier studies.

C~IV~$\lambda$1550 is also seen in absorption toward the pulsar. The line is
broad and blueshifted with a maximum velocity of  $\sim 2500 \kms$, and there
is no absorption at zero velocity. 
These are the highest velocities measured in the Crab and shows that there
exists material outside the visible nebula. This can be interpreted as 
evidence for the fast shell that has been predicted to surround the Crab
nebula (Chevalier 1977).
%For an optically thin line, the total column density of C~IV 
%is $(3.0\pm1.1)\EE{14}$ cm$^{-2}$. 
We have used a simple, spherically 
symmetric model in which the density in the shell falls off with radius
as $R^{-\eta}$ from $5\EE{18}$~cm (corresponding 
to $\approx 1680 \kms$) to derive the mass and energy of such a shell.
The conclusions from our model depend on how we tie our model into other 
observations and models. From the C~IV line alone, we find 
%for a solar abundance of carbon 
that the minimum mass and kinetic energy 
of the fast gas are $\sim 0.3 \Msun$ and $\sim 1.5\EE{49}$~ergs,
respectively. This occurs for a density slope $\eta \sim 3$. A model with 
a flat ($\eta = 0$) density profile appears unlikely as the required 
ionization structure disagrees with the modeling of Lundqvist et al. (1986). 
The maximum mass of the shell is set by progenitor models, and is unlikely 
to be much larger than $8 \Msun$. With a high shell mass, and the velocity
extending to velocities much higher than we can detect, 
the shell could carry an
energy of $10^{51}$~ergs. The signal-to-noise of C~IV~$\lambda$1550 is too 
low at high velocities to reject or confirm such a conclusion.

Adding constraints from the model of Sankrit \& Hester (1997) to those
from the C~IV line narrows down the parameter space for the shell. 
In particular, a density slope of $\eta \gtrsim 3$ is required to agree with
the interpretation of the observed [O III]-skin being a radiative shock.
%(for which the shell mass is close 
%to the upper limit of $8 \Msun$). 
For $\eta \leq 9$, the shell mass is then $\gtrsim 1.7 \Msun$ and the kinetic 
energy $\gtrsim 7\EE{49}$~ergs. Although the limit on the H$\alpha$ surface 
brightness from the search of Fesen et al. (1997) tend to favor models with 
steep density profiles, a model with $\eta = 3$ might still be possible,  
if the He/H ratio is higher than solar also in the fast shell, and the 
asymmetry of the outer shell different from that in the model of Sankrit \&
Hester (1997). 
%Given also that we are only probing the C~IV line profile along
%one line of sight, we cannot completely reject that the Crab was
%a $10^{51}$~erg event, although the combined model favors a lower energy.
%To distinguish between various models, 
%photoionization calculations and deeper H$\alpha$ searches are needed.

\acknowledgments

We thank Rob Fesen for help during preparations of the {\it HST}
observations, and 
for discussions and comments on the manuscript. We also thank Phil Plait 
for help with barycentric corrections, and Stefan Larsson for advice 
on period determination.
We thank the Swedish National Space Board, and GSFC/NASA for support which 
enabled JS and PL to visit GSFC. We are also grateful to The Swedish
Natural Science Research Council for support. JS was also supported by grants 
from the Holmberg, Hierta and Magn. Bergvall foundations. The research of 
RAC is supported through grant NAG5-8130.

\newpage

\begin{deluxetable}{lcc}
\footnotesize
\tablewidth{0pc}
\tablecaption{LOG OF STIS FUV OBSERVATIONS}
\tablehead{
\colhead{Observation}   & \colhead{Start time (MJD)}   
& \colhead{Exposure time} \nl
\multicolumn{1}{c}{{\scriptsize }}
&\multicolumn{1}{c}{{\scriptsize (51200.0+)}}
&\multicolumn{1}{c}{{\scriptsize (seconds)}}
}

\startdata

O4ZP01010  &   .54924  &  2100 \\
 
O4ZP01020   &   .60619 &  2460 \\
 
O4ZP01030    &  .67384 &  2460 \\
 
O4ZP01040    &  .74103 &  2460  \\
 
O4ZP02010 &   .81802     &   2100 \\

O4ZP02020 &   .87488     &   2460 \\

\enddata
\end{deluxetable}

%------table 2

\begin{deluxetable}{cccc}
\footnotesize
\tablewidth{0pc}
\tablecaption{LOG OF {\it NOT} OBSERVATIONS}
\tablehead{
\colhead{Grism}  & \colhead{Date} & \colhead{Wavelength range} &
\colhead{Resolution}  
%& \colhead{Exposure time}   & \colhead{Date}  & \colhead{Standard
%star} & \colhead{Position angle} 
\nl
\multicolumn{1}{c}{{\scriptsize }}
&\multicolumn{1}{c}{{\scriptsize (Dec 1998)}}
&\multicolumn{1}{c}{{\scriptsize (\AA)}}
&\multicolumn{1}{c}{{\scriptsize (\AA)}}
%&\multicolumn{1}{c}{{\scriptsize (minutes)}}
%&\multicolumn{1}{c}{{\scriptsize }}
%&\multicolumn{1}{c}{{\scriptsize (degrees)}}
}
\startdata
%CRAB-PULSAR   &    05 34 32 +22 00 53 & STIS/TIME-TA &  1425 &  2100 \\
3 & 24+25 & 3300-6400 & 6.6 \\
5 & 24+25 & 5200-9250 & 9.0 \\
6 & 27 & 3200-5500 & 4.7 \\
7 & 25 & 3820-6100 & 5.0 \\
8 & 25 & 5830-8340 & 3.6 \\

%\tablecomments{no comments}  
%\tablenotetext{1}{Days since July 14.0, 1994.}  
\enddata
\end{deluxetable}

%----------Table 3----------

\begin{deluxetable}{lccccc}
\footnotesize
\tablewidth{40pc}
\tablecaption{ABSORPTION LINES}
\tablehead{
\colhead{Species} & \colhead{$\lambda$\tablenotemark{a}}   & \colhead{$f_{\rm osc}\tablenotemark{a}$}
& \colhead{Equivalent width} & \colhead{Log (Column density)} & Log (Abundance)\nl
%\cline{2-3} \\
\multicolumn{1}{c}{{\scriptsize }}
&\multicolumn{1}{c}{{\scriptsize (\AA)}}
&\multicolumn{1}{c}{{\scriptsize }}
&\multicolumn{1}{c}{{\scriptsize (\AA)}}
&\multicolumn{1}{c}{{\scriptsize (cm$^{-2}$)}}
%&\multicolumn{1}{c}{{\scriptsize (relative to H~I)}}
}
\startdata
H~I     & 1215.67          & 0.416   & $37\pm5$ & $21.48\pm0.08$ & 12.0\\
C~I     & 1277.46          & 0.0966  & $0.18\pm0.12$ & $17.49\pm0.71$ & $7.96\pm0.76$\\
C~I     & 1329.34          & 0.0580  & $0.11\pm0.11$ & $\lesssim 18.11$ & $\lesssim 8.71$\\
C~I     & 1561.05          & 0.0804  & $0.35\pm0.17$ & $18.20\pm0.46$ & $8.70\pm0.48$\\
C~I     & 1657.59          & 0.140   & $0.54\pm0.17$ & $17.89\pm0.29$ & $8.40\pm0.30$\\
C~II    & 1335.31          & 0.128   & $0.46\pm0.24$ & $18.15\pm0.50$ & $8.65\pm0.53$\\
C~IV    & 1549.05          & 0.286   & $1.84\pm0.65$ & $14.45\pm0.16$\tablenotemark{b} & N/A\tablenotemark{b}\\
O~I\tablenotemark{c}   & 1303.49 & 0.0488 & $0.54\pm0.21$ & $18.51\pm0.36$ & $9.02\pm0.38$ \\
Mg~I    & 2852.96          & 1.830   & $0.93\pm0.33$ & $16.11\pm0.34$ & $6.62\pm0.35$\\
Mg~II   & 2796.35          & 0.612   & $0.79\pm0.25$ & $16.79\pm0.28$ & $7.30\pm0.30$\\
Mg~II   & 2803.53          & 0.305   & $0.55\pm0.27$ & $16.70\pm0.47$ & $7.20\pm0.49$\\
Al~II   & 1670.79          & 1.833   & $0.56\pm0.34$  & $16.04\pm0.57$ & $6.54\pm0.60$\\
Si~II   & 1260.42           & 1.007   & $0.60\pm0.18$ & $16.82\pm0.28$ & $7.33\pm0.29$\\
Si~II   & 1526.71          & 0.230   & $0.32\pm0.14$ & $16.57\pm0.39$ & $7.08\pm0.41$\\
Si~IV   & 1393.75          & 0.514    & $0.33\pm0.20$ & $16.63\pm0.63$ & $7.12\pm0.66$\\
Fe~II   & 2586.65          & 0.0646  & $0.27\pm0.20$ & $17.15\pm0.86$ & $7.58\pm0.96$\\
Fe~II   & 2600.17          & 0.224   & $0.40\pm0.21$ & $16.71\pm0.53$ & $7.21\pm0.56$\\

%\tablecomments{no comments}
\tablenotetext{a}{Morton (1991)}
\tablenotetext{b}{Not interstellar. The column density is calculated from ``weak-line'' theory (Morton 1991).}
\tablenotetext{c}{Could be affected by subtraction of geocoronal emission, and is probably also blended 
with Si~II~$\lambda$1304.37}

\enddata

\end{deluxetable}

%----------Table 4----------

\begin{deluxetable}{lcc}
\footnotesize
\tablewidth{0pc}
\tablecaption{ABUNDANCES}
\tablehead{
\colhead{Element} & \colhead{Solar value\tablenotemark{a}} & \colhead{This 
paper\tablenotemark{b}} \nl
%\cline{2-3} \\
%\multicolumn{1}{c}{{\scriptsize }}
%&\multicolumn{1}{c}{{\scriptsize (\AA)}}
%&\multicolumn{1}{c}{{\scriptsize }}
%&\multicolumn{1}{c}{{\scriptsize (relative to H~I)}}
}
\startdata
%C       &   8.55   &   $8.89\pm0.42$     \\
%O       &   8.87   &   $9.02\pm0.38$     \\
%Mg      &   7.58   &   $7.38\pm0.31$     \\
%Al      &   6.47   &   $6.54\pm0.60$     \\
%Si      &   7.55   &   $7.55\pm0.41$     \\
%Fe      &   7.51   &   $7.21\pm0.56$     \\
C       &   8.55   &   $8.9\pm0.4$     \\
O       &   8.87   &   $9.0\pm0.4$     \\
Mg      &   7.58   &   $7.4\pm0.3$     \\
Al      &   6.47   &   $6.5\pm0.6$     \\
Si      &   7.55   &   $7.6\pm0.4$     \\
Fe      &   7.51   &   $7.2\pm0.6$     \\

%\tablecomments{no comments}
\tablenotetext{a}{Anders \& Grevesse (1989)}
\tablenotetext{b}{Our method is likely to systematically overestimate the
abundances.
See \S 3.5.}

\enddata
\end{deluxetable}

%----------Table 5----------

\begin{deluxetable}{ccccccc}
\footnotesize
\tablewidth{40pc}
\tablecaption{PROPERTIES OF A $4 \Msun$ FAST SHELL.\tablenotemark{a}}
\tablehead{
\colhead{$\eta$\tablenotemark{b}} 
&\colhead{$n_{\rm H}(R_{\rm in})$\tablenotemark{b}} 
&\colhead{$E_{\rm sh}$}
&\colhead{$\Sigma_{{\rm H}\alpha}$\tablenotemark{c}}
&\colhead{$\Sigma_{{\rm H}\alpha}$\tablenotemark{d}}
&\colhead{$X($C~IV$)$\tablenotemark{e}} 
&\colhead{$M_{\rm trunc}$\tablenotemark{f}} \nl
%\cline{2-3} \\
\multicolumn{1}{c}{{\scriptsize }}
&\multicolumn{1}{c}{{\scriptsize ($0.714 \times \rho/{\rm m}_{\rm H}$, cm$^{-3}$)}}
&\multicolumn{1}{c}{{\scriptsize ($10^{51}$ ergs)}}
&\multicolumn{1}{c}{{\scriptsize (ergs~cm$^{-2}$~s$^{-1}$~sr$^{-1}$)}}
&\multicolumn{1}{c}{{\scriptsize (ergs~cm$^{-2}$~s$^{-1}$~sr$^{-1}$)}}
&\multicolumn{1}{c}{{\scriptsize }}
&\multicolumn{1}{c}{{\scriptsize ($\Msun$)}}
}
\startdata
0 &  0.12  &  0.99 (0.03)\tablenotemark{g}  &  $9.3(4.1)\tablenotemark{g}~\EE{-9}$  & $9.2(3.9)\tablenotemark{g}~\EE{-9}$ & 1\tablenotemark{h} & 0.43 \\
1 &  0.32  &  0.86 (0.05) &  $2.4(1.8)\EE{-8}$  & $2.1(1.6)\EE{-8}$ &  0.62   & 0.78 \\
2 &  0.77  &  0.72 (0.08) &  $8.1(7.6)\EE{-8}$  & $6.1(5.6)\EE{-8}$ &  0.28   & 1.29 \\
3 &  1.62  &  0.56 (0.11) &  $2.7(2.7)\EE{-7}$  & $1.7(1.6)\EE{-7}$ &  0.14   & 1.92 \\
4 &  2.94  &  0.42 (0.14) &  $7.4(7.4)\EE{-7}$  & $3.8(3.8)\EE{-7}$ &  0.081  & 2.57 \\
5 &  4.66  &  0.32 (0.15) &  $1.6(1.6)\EE{-6}$  & $6.9(6.9)\EE{-7}$ &  0.053  & 3.11 \\
7 &  8.71  &  0.21 (0.16) &  $4.7(4.7)\EE{-6}$  & $1.4(1.4)\EE{-6}$ &  0.030  & 3.71 \\
9 &  13.0  &  0.17 (0.15) &  $9.1(9.1)\EE{-6}$  & $1.8(1.8)\EE{-6}$ &  0.021  & 3.92 \\

%\tablecomments{no comments}
\tablenotetext{a}{Minimum and maximum radii are $5\EE{18}$~cm and $1.9\EE{19}$~cm, respectively, corresponding to the velocities $\approx 1680 \kms$ and $\approx 6370 \kms$.}
\tablenotetext{b}{$n_{\rm H}(R) = n_{\rm H}(R_{\rm in})~(R/R_{\rm in})^{- \eta}$}
\tablenotetext{c}{H$\alpha$ surface brightness at impact parameter $p = R_{\rm in}$, i.e., just at the edge of the observed nebula. ($T = 2\EE4$ K).}
\tablenotetext{d}{H$\alpha$ surface brightness at impact parameter $p = 1.1 R_{\rm in}$, i.e., $\sim 17\arcsec$ outside the observed nebula. ($T = 2\EE4$ K).}
\tablenotetext{e}{Fraction of carbon in C IV to obtain the observed optical depth in C~IV~$\lambda1550$ at $\sim 1680 \kms$. ($X($C$)/X($H$) = 3.5\EE{-4}$).}
\tablenotetext{f}{How much of the $4\Msun$ shell that resides inside $3000 \kms$. Density and $X($C~IV$)$ at $R_{\rm in}$ are assumed to be the same as for the $4 \Msun$ shell with maximum velocity $6370 \kms$.}
%\tablenotetext{f}{Mass of the shell if gas with velocity $> 3000 \kms$ is absent. Density and $X($C~IV$)$ at $R_{\rm in}$ are the same as for the $4 \Msun$ shell with maximum velocity $6370 \kms$.}
\tablenotetext{g}{Values in parentheses are for the case with the shell only reaching $3000 \kms$, i.e., for the masses in the last column in the table.}
\tablenotetext{h}{For $\eta = 0$ the maximum optical depth in C~IV~$\lambda1550$ is too small even with $X($C~IV$) = 1$. (See Fig. 10).}

\enddata
\end{deluxetable}

\newpage
\clearpage

\begin{figure*} \centering \vspace{10.0 cm}
\includegraphics{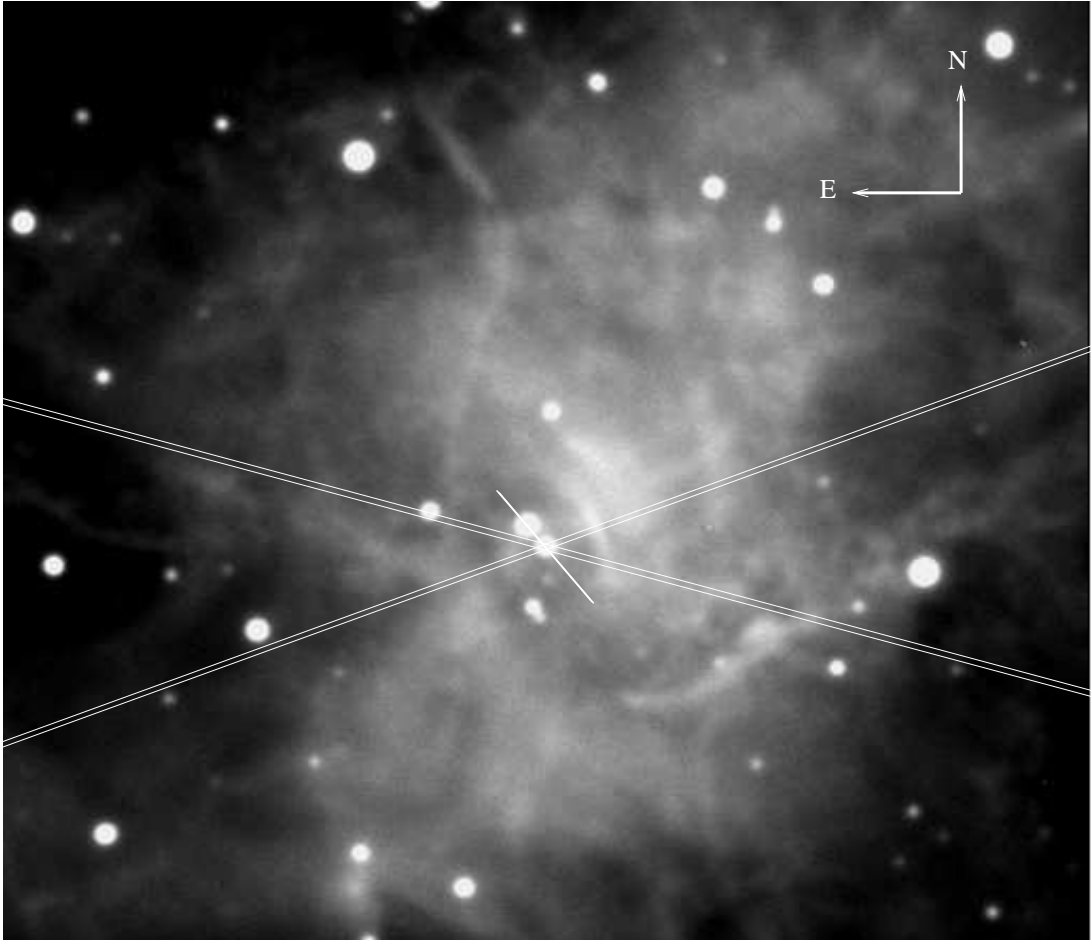}
\caption{
An $R$-band image of the Crab nebula obtained at 
the {\it NOT} in December, 1998. 
Shown are the extents and position angles of the slits 
used in the optical and in the FUV. The STIS-FUV slit is 25\arcsec~long.
The NUV observation (G98) used 
a 2\arcsec$\times$2\arcsec aperture centered on the pulsar.
}
\label{ima}
\end{figure*}

\clearpage

\begin{figure*} \centering \vspace{8.0 cm}
\includegraphics{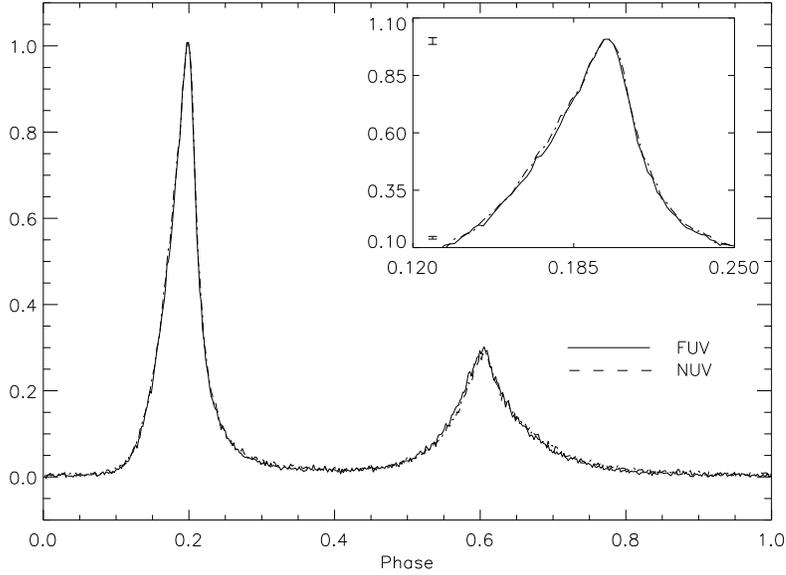}
\caption{
Pulse profile of the Crab pulsar in the FUV (full line). 
Also shown is the NUV pulse profile from G98 (dashed line). 
The FUV and NUV data were processed in the same way. 
The blow-up of the primary peak shows the FUV to be slightly narrower.
The statistical errors range from 0.003 in the valleys to 0.014 at the peak 
for both the FUV and NUV profiles, as illustrated by the 
($\pm1\sigma$) error bars in the insert. 
Note that the count rate of the pulsar signal is close to zero in the 
interpulse region.
}
\label{pulse}
\end{figure*}

\begin{figure*} \centering \vspace{8.0 cm}
\includegraphics{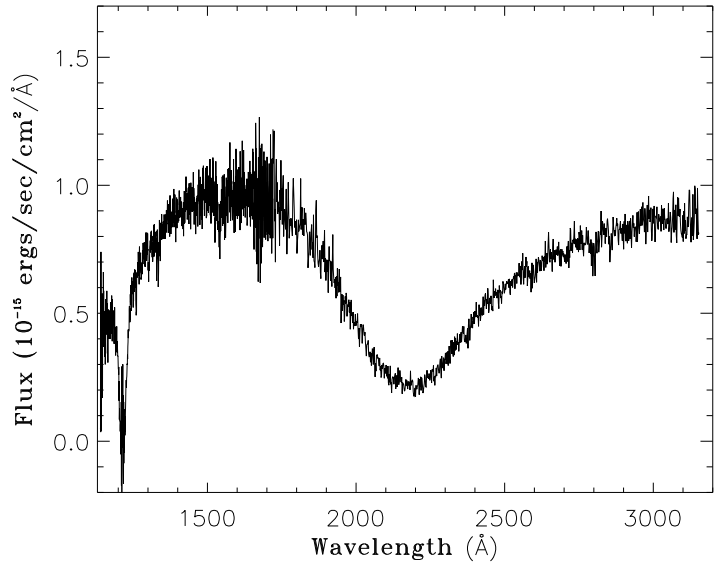}
\caption{ 
Spectrum of the Crab pulsar in the UV. This is a combination of our
FUV ($1140-1730$~\AA) observations obtained in January 1999, and the NUV 
($1600-3200$~\AA) data from G98. The spectra overlap nicely. 
The ${\rm Ly}\alpha$ absorption dip shows some residuals from the
subtraction of strong geocoronal emission.
}
\label{uv}
\end{figure*}

\clearpage

\begin{figure*} \centering \vspace{10.0 cm}
\includegraphics{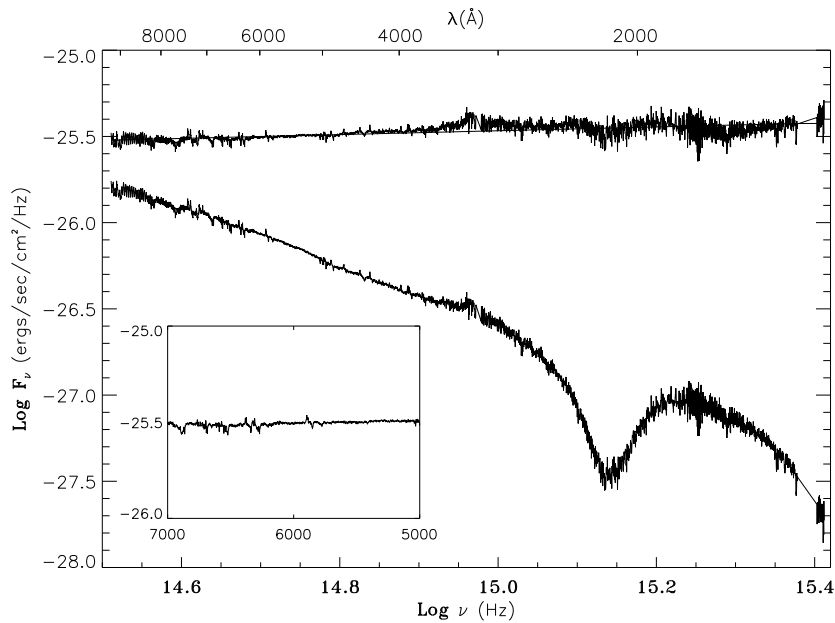}
\caption{ 
Spectrum of the Crab pulsar in the UV and in the optical. The optical
data are from
{\it NOT}. The lower spectrum shows the flux-calibrated spectrum without 
dereddening. The optical spectrum connects rather well to the NUV data. 
%although the bluest part is rather uncertain. 
The uppermost spectrum has 
been dereddened with $E(B-V)=0.52$ and $R=3.1$. 
The full line shows the best power law fit, which has
the spectral index $\alpha_{\nu}=0.11$.
The insert shows a blowup of the optical
region where Nasuti et al. (1996) reported a broad absorption dip in the 
spectrum.
}
\label{pulse}
\end{figure*}

\begin{figure*} \centering \vspace{8.0 cm}
\includegraphics{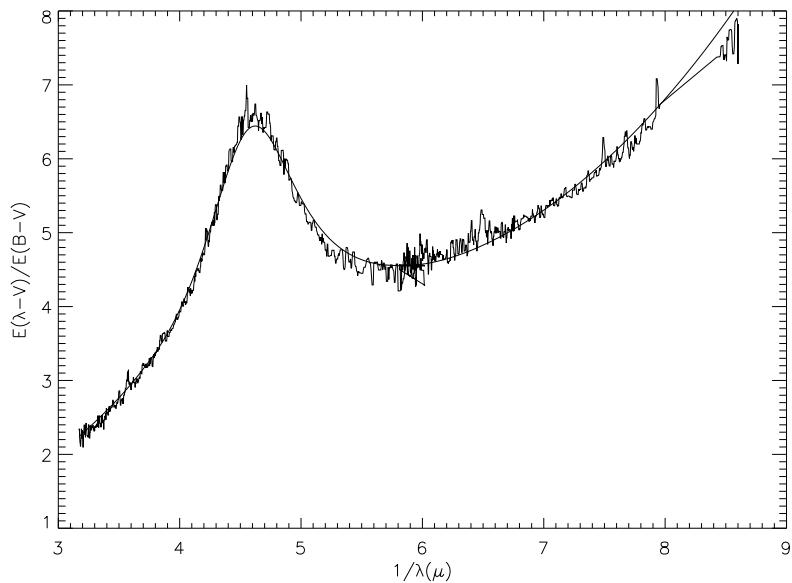}
\caption{
The UV extinction curve in the direction toward the Crab. This curve
was obtained by assuming the pulsar intrinsic spectrum to follow a power law.
For comparison, the mean galactic extinction curve from Fitzpatrick (1999) is
also shown.
}
\label{redcurve}
\end{figure*} 

\clearpage

\begin{figure*} \centering \vspace{10.0 cm}
\includegraphics{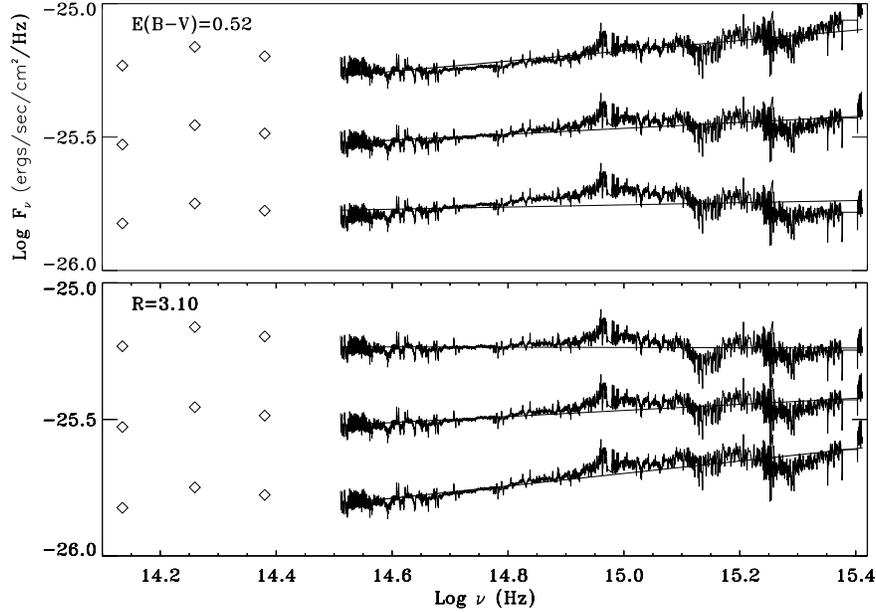}
\caption{
The Crab pulsar spectrum dereddened with different values of $E(B-V)$ and $R$.
The upper panel shows the dereddened spectrum for $E(B-V)=0.52$ and three
different values of $R$: 2.9 (upper), 3.1 and 3.3 (lower). The IR data points
are from Eikenberry et al. (1997) and are not included in the power law fits.
The fitted power laws have $\alpha_{\nu}=$
0.19, 0.11 and 0.041, respectively.
The lower panel shows dereddening for $R=3.1$ and $E(B-V)=$0.49 (upper), 0.52
and 0.55 (lower). The fitted power laws have spectral 
indices $\alpha_{\nu}= -0.005, 0.11$ and $0.23$, respectively. 
The individual spectra in both panels have been shifted by $-0.3$, 0.0, 
and +0.3 dex in the vertical direction for clarity.
}
\label{reddall}
\end{figure*}

\begin{figure*} \centering \vspace{8.0 cm}
\includegraphics{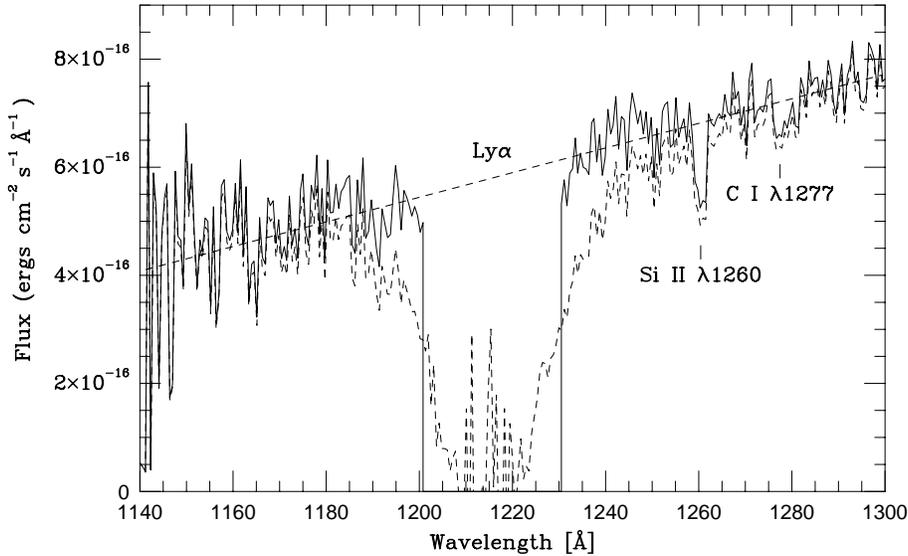}
\caption{
FUV spectrum around Ly$\alpha$. The dashed line shows the original
data, and the solid line shows the spectrum after compensation for
absorption in the damping wings of Ly$\alpha$. The column density of neutral
hydrogen toward the Crab determined in this way 
is $N$(H~I) = $(3.0\pm0.5)\EE{21}$ cm$^{-2}$. The straight dashed line shows 
a continuum fit to the spectrum after correction for Ly$\alpha$ absorption.
}
\label{lyalpha}
\end{figure*} 

\clearpage

\begin{figure*} \centering \vspace{15.0 cm}
\includegraphics{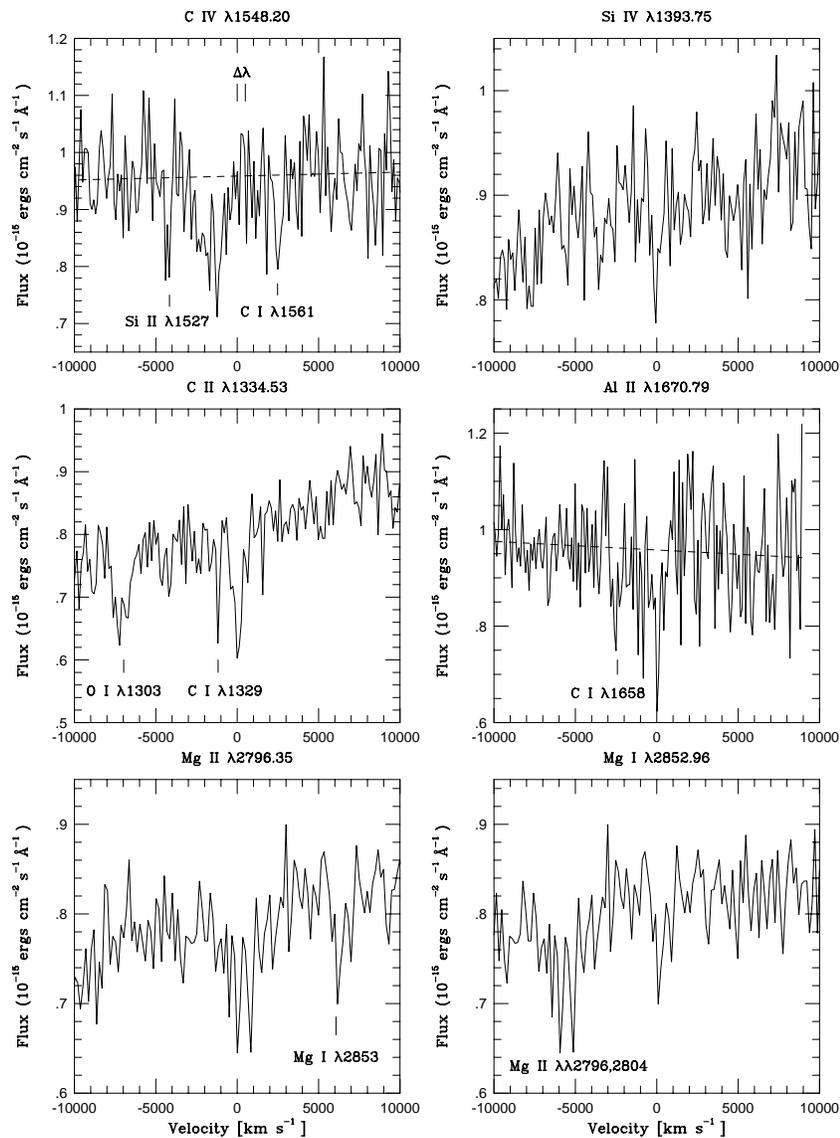}
\caption{
Absorption lines seen in the spectrum toward the Crab pulsar. Only
C~IV~$\lambda$1550 shows absorption which cannot be interstellar (see Fig.
10). The difference in velocity between the two components of the C~IV
doublet is shown by the two vertical lines marked by ``$\Delta\lambda$''.
The equivalent widths of all lines are given in Table 3. Note that the
zero velocity of the C~II and C~IV multiplets in the figure corresponds to
the most blueward component of the multiplets, while Table 3 lists 
weighted wavelengths.
}
\label{abslines}
\end{figure*} 

\clearpage

\begin{figure*} \centering \vspace{8.0 cm}
\includegraphics{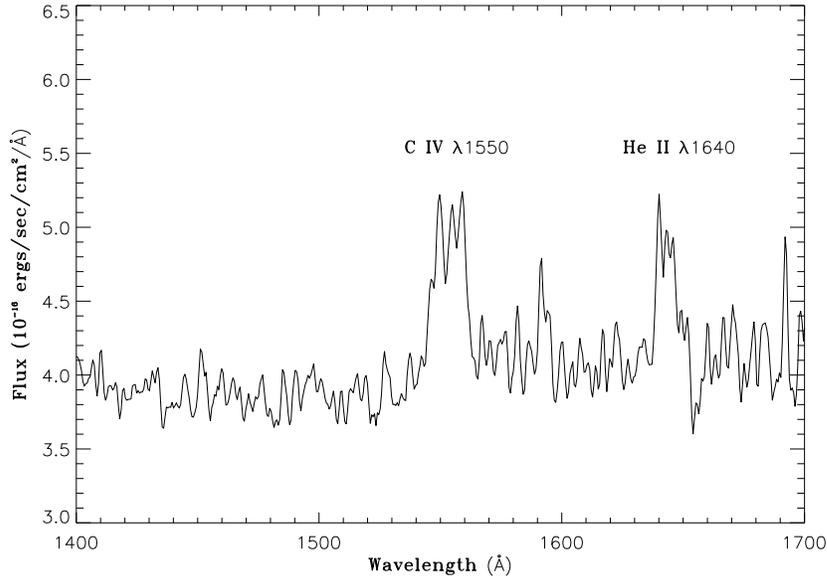}
\caption{
FUV emission from the Crab nebula in the region around the pulsar. The 
spectrum was obtained by averaging the emission from two 8\farcs2 long
region above and below 
the pulsar position along the 0\farcs5 slit. The continuum has been flattened 
out and the spectrum smoothed with a 3 pixel boxcar average.
}
\label{pulse}
\end{figure*}

\begin{figure*} \centering \vspace{8.0 cm}
\includegraphics{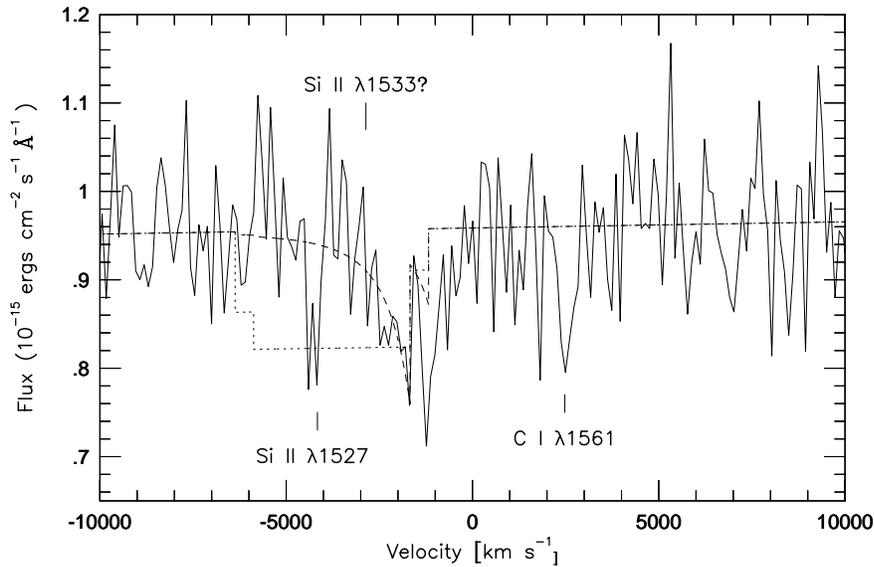}
\caption{
Pulsar spectrum around C~IV~$\lambda1550$. No absorption at zero velocity
of C~IV~$\lambda1550$ is seen. Instead the line is blueshifted, reaching a
maximum velocity $\sim 2500 \kms$. The spectrum also shows the interstellar
lines Si~II~$\lambda$1527 and C~I~$\lambda$1561, neither of which blends with
the C~IV line. Si~II~$\lambda$1533, which sometimes accompanies 
Si~II~$\lambda$1527 in interstellar spectra, appears to be absent, and cannot
explain the observed absorption at velocities $\lesssim 2700 \kms$. Overlaid 
on the observed spectrum are two of the models for the fast shell in Table 
5: $\eta = 0$ (dotted) and $\eta = 3$ (dashed), where $\eta$ is defined 
from $\rho(R) = \rho(R_{\rm in})~(R/R_{\rm in})^{-\eta}$, and $R_{\rm in}$
is the inner radius of the shell. We have assumed 
that $R_{\rm in}=5\EE{18}$~cm, which corresponds to a minimum shell velocity
of $\approx 1680 \kms$. See Table 5 for details of these and other models of
the shell.
}
\label{CIVline}
\end{figure*}

\end{document}